\documentclass[fleqn,usenatbib]{mnras}

\usepackage{comment}
\usepackage{newtxtext,newtxmath}

\usepackage[T1]{fontenc}

\DeclareRobustCommand{\VAN}[3]{#2}
\let\VANthebibliography\thebibliography
\def\thebibliography{\DeclareRobustCommand{\VAN}[3]{##3}\VANthebibliography}

\usepackage{kotex}
\usepackage{color}
\definecolor{Dred}{rgb}{0.312,0.070,0.070}
\definecolor{Dblue}{rgb}{0.070,0.070,0.312}
\definecolor{Dgreen}{rgb}{0.070,0.312,0.070}
\definecolor{Db}{rgb}    {0.050,0.0,0.320}

\newcounter{note}
\let\oldmarginpar\marginpar
\renewcommand\marginpar[1]{\-\oldmarginpar[\raggedleft\footnotesize #1]{\raggedright\footnotesize #1}}

\usepackage{graphicx}	
\usepackage{amsmath}	
\usepackage{caption}
\usepackage{subcaption}
\usepackage{threeparttable}

\title[Magnetic field Strengths of the CTA 102 Jet]{Magnetic Field Strengths of the Synchrotron Self-Absorption Region in the Jet of CTA~102 During Radio Flares}   

\author[S.-H. Kim et al.]{
Sang-Hyun Kim,$^{1,2}$
Sang-Sung Lee,$^{1,2}$\thanks{Corresponding author: Sang-Sung Lee (\href{mailto:sslee@kasi.re.kr}{sslee@kasi.re.kr})}
Jee Won Lee,$^{1}$
Jeffrey A. Hodgson,$^{1,3}$
Sincheol Kang,$^{1,2}$
\newauthor{Juan-Carlos Algaba,$^{4}$
Jae-Young Kim,$^{1,5}$
Mark Hodges,$^{6}$
Ivan Agudo,$^{7}$
Antonio Fuentes, $^{7}$
Juan Escudero,$^{7}$}
\newauthor{Ioannis Myserlis,$^{8}$
Efthalia Traianou,$^{5}$
Anne L\"{a}hteenm\"{a}ki,$^{9,10}$
Merja Tornikoski,$^{9}$
Joni Tammi,$^{9}$}
\newauthor{Venkatessh Ramakrishnan$^{9,11}$
and Emilia J\"{a}rvel\"{a}$^{9,12}$}
\\
$^{1}$Korea Astronomy and Space Science Institute, 776 Daedeok-daero, Yuseong-gu, Daejeon 34055, Korea\\
$^{2}$University of Science and Technology, Korea, 217 Gajeong-ro, Yuseong-gu, Daejeon 34113, Korea\\
$^{3}$Department of Physics and Astronomy, Sejong University, 209 Neungdong-ro, Gwangjin-gu, Seoul, South Korea\\
$^{4}$Department of Physics, Faculty of Science, University of Malaya, 50603 Kuala Lumpur, Malaysia\\
$^{5}$Max Planck Institute for Radio Astronomy, Auf dem H\"{u}gel, 69
D-53121 Bonn, Germany\\
$^{6}$Owens Valley Radio Observatory, California Institute of Technology, Pasadena, CA 91125, USA\\
$^{7}$Instituto de Astrof\'{i}sica de Andaluc\'{i}a-CSIC, Glorieta de la Astronom\'{i}a s/n, E—18008, Granada, Spain\\
$^{8}$Instituto de Radioastronom\'{i}a Milim\'{e}trica, Avenida Divina Pastora, 7, N\'{u}cleo Central, E—18012, Granada, España\\
$^{9}$Aalto University Mets\"{a}hovi Radio Observatory, Mets\"{a}hovintie 114, 02540 Kylm\"{a}l\"{a}, Finland\\
$^{10}$Aalto University Department of Electronics and Nanoengineering,
P.O. BOX 15500, FI-00076 AALTO, Finland\\
$^{11}$Astronomy Department, Universidad de Concepción, Casilla 160-C, Concepción, Chile\\
$^{12}$European Space Agency, European Space Astronomy Centre, C/ Bajo el Castillo s/n, 28692 Villanueva de la Cañada, Madrid, Spain\\
}

\date{Accepted 2021 November 25. Received 2021 November 25; in original form 2021 May 12}

\pubyear{2021}

\begin{document}
\label{firstpage}
\pagerange{\pageref{firstpage}--\pageref{lastpage}}
\maketitle

\begin{abstract}
CTA 102 is a blazar implying that its relativistic jet points towards Earth and emits synchrotron radiation produced by energetic particles gyrating in the magnetic field.
This study aims to figure out the physical origins of radio flares in the jet, including the connection between the magnetic field and the radio flares.
The dataset in the range 2.6--343.5~GHz was collected over a period of $\sim$5.5~years (2012 November 20--2018 September 23).
During the data collection period, seven flares at 15~GHz with a range of the variability time-scale of roughly 26--171~days were detected.
The quasi-simultaneous radio data were used to investigate the synchrotron spectrum of the source.
We found that the synchrotron radiation is self-absorbed.
The turnover frequency and the peak flux density of the synchrotron self-absorption (SSA) spectra are in the ranges of $\sim$42--167~GHz and $\sim$0.9--10.2~Jy, respectively.
From the SSA spectra, we derived the SSA magnetic field strengths to be $\sim$9.20~mG, $\sim$12.28~mG, and $\sim$50.97~mG on 2013 December 24, 2014 February 28, and 2018 January 13, respectively.
We also derived the equipartition magnetic field strengths to be in the range $\sim$24--109~mG.
The equipartition magnetic field strengths are larger than the SSA magnetic field strengths in most cases, which indicates that particle energy mainly dominates in the jet.
Our results suggest that the flares in the jet of CTA~102 originated due to particle acceleration.
We propose the possible mechanisms of particle acceleration.
\end{abstract}




\section{Introduction} \label{sec:intro}

Blazars are a subclass of active galactic nuclei (AGN) and are among the most powerful objects in the universe.
It is believed that supermassive black holes (SMBHs) with a mass range of $\sim$$10^6$--$10^{10}~\mathrm{M_\odot}$ are a critical part of the central engine of blazars \citep{Blandford+2019}.
Relativistic jets are the luminous and collimated outflows emitting synchrotron radiation resulting from accelerated electrons gyrating magnetic field lines within those jets \citep{Boettcher+2012}.
Magnetic fields, anchored in the accretion disc or the ergosphere of SMBHs, make a crucial contribution to the formation, collimation, and acceleration of these relativistic jets \citep{Blandford&Znajek+1977, Blandford&Payne+1982}.

Blazars are divided into two subclasses, flat-spectrum radio quasars (FSRQs) and BL Lac objects (BL Lacs).
Both subclasses of blazars are characterized by relativistically boosted emission, extremely variable emissions extending from radio to $\gamma$-ray energies, and superluminal jet motions \citep{Jorstad+2005, Marscher+2008, Hovatta+2008}.
These features can mostly be attributed to the small viewing angles \citep[e.g., $<10$~deg;][]{Hovatta+2009} from our line of sight \citep[leading to the relativistic effects;][]{Urry&Padovani+1995}.
These effects can be quantified by the Doppler factor ($\delta$), which is defined as $\delta=[\Gamma(1-\beta\cos{\theta})]^{-1}$, where $\Gamma$ is the bulk Lorentz factor, $\beta$ is the intrinsic velocity of the jet in units of the speed of light, and $\theta$ is the viewing angle.
Since it is difficult to directly estimate $\beta$ and $\theta$, there is some difficulty in obtaining reliable estimates of the Doppler factor.
Alternatively, an indirect method has been proposed to estimate $\delta$ using the observed variability \citep{Lahteenmaki&Valtaoja+1999, Hovatta+2009, Liodakis+2017}.
Therefore, characterizing the Doppler factor significantly contributes to the understanding of the highly variable emissions seen from blazars.

CTA~102 (also known as J2232+114) is a FSRQ with redshift $z = 1.037$ \citep{Schmidt+1965}.
Its luminosity distance is $D_\mathrm{L} = 6943$~Mpc (adopting a cosmology with $\Omega_\mathrm{m}=0.27$, $\Omega_\mathrm{\Lambda}=0.73$, and $H_0=71$~$\mathrm{km \cdot s^{-1} \cdot {Mpc}^{-1}}$).
It harbors a SMBH with a mass of $8.5\times10^8~\mathrm{M_\odot}$ \citep{Zamaninasab+2014}.
As the source is a blazar, the strong emitted electromagnetic radiation ranging from radio to $\gamma$-rays shows rapid variability, especially in radio \citep{Rantakyro+2003, Lister+2009b, Fromm+2011, Casadio+2015}.
The 2006 radio flare from CTA~102, observed at centimeter (cm) to millimeter (mm) wavelengths, was investigated by \citet{Fromm+2011}.
The behavior of the observed radio flux enhancements is in agreement with the predictions from the shock-in-jet model, where the outburst originates from a shock wave passing through the relativistic jet \citep{Marscher&Gear+1985}.
An mm-wave radio outburst was reported in 2012 by \citep{Casadio+2015}.
Single-dish observations at cm- to mm-wavelengths were reported to have reached a peak at the end of 2016 \citep{Raiteri+2017, D'Ammando+2019}.
Very long baseline interferometry (VLBI) observations of blazars typically show a compact bright region at the upper end of the jet which is normally referred to as the VLBI core \citep{Jorstad+2005, Lister+2009b}.
VLBI observations at $43~\mathrm{GHz}$ showed that the flux density of the VLBI core peaked during the same epoch as when the single-dish measurements reached their maximum \citep{Casadio+2019}.
Polarimetric multi-frequency observations consistently suggest the presence of a helical magnetic field in the jet of CTA~102 \citep{Hovatta+2012, Casadio+2015, Raiteri+2017, Li+2018, Park+2018, Casadio+2019}, suggesting that the magnetic field may play an important role in the radio emissions from the jet.

Many studies have estimated magnetic field strengths in blazars usually using one of four different methods based on: 1) the broadband spectral energy distribution (SED) model, 2) the core-shift effect, 3) the synchrotron luminosity, and 4) the synchrotron self-absorption (SSA) spectrum.
We will discuss how each method estimates the magnetic field strengths in blazars in the following.
The first method is based on a blazar spectrum typically exhibiting two humps; one at lower energies and one at higher energies.
The lower energy hump is thought to originate from synchrotron emission, and the higher energy hump is due to the scattering of photons off relativistic electrons; these photons come from synchrotron emission or AGN components (e.g., accretion disk, broad-line region, or dusty torus).
From this basis, the magnetic field strength can be estimated \citep{Mastichiadis&Kirk+1997, Tramacere+2009, Tavecchio&Ghisellini+2014}.
The magnetic field strengths of CTA~102 were found to be constrained in the range 0.1--6.1~G assuming a leptonic model \citep{Zacharias+2017, Raiteri+2017, Gasparyan+2018, Prince+2018}, or are in the range 50--80~G assuming a hadronic model \citep{Zacharias+2019}.
In the second method, the core-shift effect, the frequency-dependent position shift of the VLBI core can also be used, under several assumptions, to estimate magnetic field strengths at 1~pc from the base of the jet \citep{Lobanov+1998, O'Sullivan&Gabuzda+2009, Algaba+2012, Pushkarev+2012, Fromm+2013b, Li+2018}.
The magnetic field strengths of CTA~102 at 1~pc from the jet base using core-shift measurements were estimated to be 0.2--2.12~G \citep{Algaba+2012, Pushkarev+2012, Fromm+2013b, Li+2018}.
For the third method, \citet{LeeSS+2016a} estimated the synchrotron luminosities of 109 sources including CTA~102 using the VLBI core fluxes at 2, 8, 15, and 86~GHz that were obtained between 1993 and 2003.
From these synchrotron luminosities, they inferred the magnetic field strengths at 1~pc from the jet apex based on the standard jet model \citep{Blandford&Konigl+1979}, this is an idealized model of a steady radio jet that assumes a conical geometry with a narrow opening angle, where the magnetic field in the jet that becomes weaker as farther away from the jet apex, and assumes approximate equipartition between the electron energy density and the magnetic field energy density.
In the fourth and final method, the magnetic field strengths are computed from the SSA spectrum by assuming a uniform, spherical, and optically thick synchrotron source \citep{Marscher+1983, Hirotani+2005}.
Combining the spectral and geometrical information of the optically thick synchrotron source, the magnetic field strengths in other blazars have been estimated \citep{Rani+2013, Hodgson+2017, LeeJW+2017, Algaba+2018b}.

In this work, we present the results from a study on the physical origins of radio flares from CTA~102 during the period from 2012 to 2018.
The magnetic field strengths from the synchrotron spectrum were estimated using multi-frequency data.
We investigated a connection between radio flares and magnetic field strengths.
In Section \ref{sec:obs}, we describe details about the observations and the multi-frequency data.
Our main results are explained in Section \ref{sec:results}.
An interpretation of our results is presented in Section \ref{sec:discussion}.
In Section \ref{sec:conclusions}, we summarize the results and discuss our conclusions.

\begin{table*}
\caption{Summary of the multi-frequency observations of CTA~102}
\label{table: all observations}
\centering
{
\begin{threeparttable}
\begin{tabular}{cccccc}
\hline
Observatory                         & $\nu_\mathrm{obs}$ (GHz) & Range of $S_\nu$ (Jy) & $t_{\rm int}$ (days) & Date                  & MJD          \\
(1)                                 & (2)                     & (3)                  & (4)                 & (5)                   & (6)          \\
\hline
OVRO 40-m                           & 15                      & 2.85--3.99           & 7                   & 2012/11/20--2018/7/21 & 56251--58320 \\
Mets\"{a}hovi 14-m & 37                      & 2.14--6.30           & 5                   & 2012/12/5--2018/5/29  & 56266--58267 \\
KVN 21-m                            & 43                      & 2.38--6.95           & 42                  & 2013/3/28--2018/5/31  & 56379--58270 \\
                                    & 22                      & 2.86--4.84           & 43                  & 2013/3/28--2018/5/31   & 56379--58270 \\
IRAM 30-m                           & 230                     & 0.96--8.94           & 51                  & 2013/4/30--2018/6/7   & 56412--58276 \\
                                    & 86                      & 2.02--10.52          & 33                  & 2013/4/30--2018/6/19  & 56412--58288 \\
SMA                                 & 230                     & 0.98--8.26           & 23                  & 2013/2/5--2018/9/23   & 56328--58384 \\
ALMA                                & 343.5                   & 0.62--5.82           & 21                  & 2013/10/02--2017/9/9  & 56567--58005 \\
                                    & 233                     & 1.61--7.00           & 70                  & 2014/10/18--2017/3/29 & 56948--57841 \\
                                    & 103.5                   & 1.58--7.16           & 14                  & 2013/6/11--2017/9/29  & 56454--58025 \\
                                    & 91.5                    & 1.66--7.11           & 13                  & 2013/6/11--2017/9/29  & 56454--58025 \\
Effelsberg 100-m                    & 43                      & 1.29--4.92           & 36                  & 2012/12/1--2014/12/2  & 56262--56993 \\
                                    & 32                      & 2.57--4.56           & 29                  & 2012/12/1--2014/3/3   & 56262--56719 \\
                                    & 23.05                   & 2.20--4.09           & 33                  & 2012/12/1--2014/12/2  & 56262--56993 \\
                                    & 14.60                   & 2.90--3.95           & 33                  & 2012/12/1--2014/12/2  & 56262--56993 \\
                                    & 10.45                   & 3.21--4.21           & 29                  & 2012/12/2--2014/12/2  & 56263--56993 \\
                                    & 8.35                    & 3.41--4.17           & 29                  & 2012/12/2--2014/12/2  & 56263--56993 \\
                                    & 4.85                    & 4.26--4.72           & 31                  & 2012/12/2--2014/12/2  & 56263--56993 \\
                                    & 2.64                    & 5.59-6.15            & 36                  & 2012/12/2--2014/12/2  & 56263--56993 \\
\hline    
\end{tabular}
\begin{tablenotes}
\item Note. Column designation: (1) observatory name, (2) observing frequency in GHz, (3) range of the flux density in Jy, (4) mean cadence of data points in days, (5) observing date in year/month/day, and (6) observing date in modified Julian date.
\end{tablenotes}
\end{threeparttable}
}
\end{table*}

\section{Observations and Data Acquisition}  \label{sec:obs}
Table \ref{table: all observations} summarizes the multi-frequency observations of CTA~102.
In this section, we describe the observations made by the single-dish radio telescopes in the Korean VLBI Network (KVN) and describe how we reduced the data to obtain the flux density of the source at 22 and 43~GHz in Section \ref{subsec:KVN}.
Information on multi-frequency data obtained by other radio facilities is described in Section \ref{subsec:mf data}.

\subsection{KVN Single-Dish Observations}    \label{subsec:KVN}
We observed CTA~102 at 22 and 43~GHz using the 21-m radio telescopes of the KVN: KVN Yonsei (KY), KVN Ulsan (KU), and KVN Tamna (KT), from March 2013 to May 2018 (MJD 56379--58270).
The observations were performed in cross scan mode, in other words, a one-dimensional on-the-fly method that uses two sub-scans in the azimuth direction and two in the elevation direction \citep[e.g.,][]{LeeJA+2017}.
These observations were made as part of the KVN key science program for interferometric MOnitoring of GAmma-ray Bright AGNs \citep[iMOGABA;][]{LeeSS+2016b}.

The cross scan observation mode enables us to obtain the single-dish flux density of the source.
Each sub-scan provides a profile of the amplitude expressed in terms of antenna temperature $T^{\ast}_\mathrm{A}$, as a function of the relative position angle with respect to the pointing center.
For calibrating the cross-scan data, we used the GILDAS package in the CLASS software\footnote{\url{https://www.iram.fr/IRAMFR/GILDAS}}.
We fitted a linear function to the baseline from each amplitude profile, to remove background emission from the instruments and sky.
Then, we fitted a Gaussian curve to the amplitude and obtained its best-fitting parameters: the peak intensity in Kelvin, the integrated area in arcsecond, the peak position offset in arcsecond, and the width in arcsecond.
We did not make use of data if their Gaussian-fitted positions offsets they produced were larger than 30~arcsec and if the profile widths were wider than 20~{per cent} of the beam size at each frequency.
The uncertainty in the peak intensity was estimated by propagating errors in the area and the width.

The peak intensity, $T^{\ast}_\mathrm{A}$, is converted into flux density $S_\mathrm{SD}$ using the following equation:
\begin{equation}    \label{eq:S_SD}
    S_\mathrm{SD} = \frac{2k_\mathrm{B}T^{\ast}_\mathrm{A}}{ \eta_\mathrm{A} A_\mathrm{geom}}~[\mathrm{Jy}], \\
\end{equation}
where $k_\mathrm{B}$ is the Boltzmann constant, $\eta_\mathrm{A}$ is the aperture efficiency, and $A_\mathrm{geom}$ is the geometric area of the telescope.
If the source was observed at an elevation between 40 and 70~deg, the normalized gain $G_\mathrm{norm}$ would be close to the maximum gain for the KVN \citep[e.g., $\sim$95--99~{per cent};][]{LeeSS+2011}.
The KVN observatory provides varying antenna aperture efficiencies on a half-yearly basis using regular gain measurements\footnote{\url{https://radio.kasi.re.kr/kvn/main_kvn.php}}.
We referred to these gain values when applying Eq. \ref{eq:S_SD} to translate the $T^{\ast}_\mathrm{A}$ to $S_\mathrm{SD}$.
We derived aperture efficiencies from the single-dish polarization observations of Jupiter and have included these in our calculations.
If the gain values were not available for particular epochs, we derived $G_\mathrm{norm}$ which has a second-order polynomial form: $G_\mathrm{norm} = A0 \times El^2 + A1 \times El + A2$, where $El$ is the elevation in deg, and $A0$, $A1$, and $A2$ are the fitted parameters which are presented on the KVN website \citep{LeeSS+2011}.
The aperture efficiency could then obtained using $\eta_\mathrm{A} = \eta_\mathrm{A, max} \times G_\mathrm{norm}$.
The flux density $S_\mathrm{SD}$ was corrected for antenna pointing offsets (peak position offsets) in both azimuthal and elevational directions.
Then, we regard the mean of the corrected flux densities as the true flux density.
If multiple scans were available for particular epochs, the mean and standard deviation values are used for the flux densities and their statistical errors, respectively.
For epochs with only a single scan available, we propagated errors from the fit results to estimate the total uncertainty.
Since the KVN has three telescopes, we used the inverse variance weighted mean value of the flux densities from those telescopes to minimize the statistical error and obtain the most reliable measurement.
The results for the single-dish flux density are summarized in Table \ref{table: KVN SD fluxes}.

\begin{table}
\caption{KVN single-dish fluxes} 
\label{table: KVN SD fluxes}
\centering
{
\begin{threeparttable}
\begin{tabular}{cccccccc}
\hline
Date & MJD & $S_\mathrm{SD,22}$ (Jy) & $S_\mathrm{SD,43}$ (Jy) \\
(1) & (2) & (3) & (4) \\
\hline
2013/3/28                     & 56379 & $4.15\pm0.02$ & $4.14\pm0.06$ \\
2013/4/11                     & 56393 & $3.97\pm0.06$ & $4.09\pm0.04$ \\
2013/5/7                      & 56419 & $3.88\pm0.07$ & $3.46\pm0.13$ \\
2013/9/24                     & 56559 & $3.40\pm0.10$ & $2.86\pm0.24$ \\
2013/10/15                    & 56580 & $3.23\pm0.07$ & $2.73\pm0.02$ \\
2013/11/19                    & 56615 & $3.01\pm0.06$ & $2.80\pm0.07$ \\
2013/12/24                    & 56650 & $3.14\pm0.04$ & $3.16\pm0.05$ \\
2014/1/27                     & 56684 & $3.40\pm0.05$ & $3.28\pm0.09$ \\
2014/2/28                     & 56716 & $3.46\pm0.07$ & $3.21\pm0.22$ \\
2014/3/22                     & 56738 & $3.45\pm0.09$ & $3.11\pm0.09$ \\
2014/4/22                     & 56769 & $3.26\pm0.06$ & $2.45\pm0.10$ \\
2014/6/13                     & 56822 & $3.17\pm0.10$ & $2.38\pm0.14$ \\
2014/9/1                      & 56902 & $3.00\pm0.17$ & $2.42\pm0.17$ \\
2014/9/27                     & 56928 & $2.86\pm0.13$ & $2.79\pm0.21$ \\
2014/10/29                    & 56960 & $2.90\pm0.13$ & $2.75\pm0.10$ \\
2014/11/28                    & 56990 & $3.08\pm0.28$ & $3.01\pm0.48$ \\
2014/12/25                    & 57017 & $3.24\pm0.05$ & $2.70\pm0.08$ \\
2015/1/15                     & 57038 & $3.27\pm0.07$ & $2.54\pm0.12$ \\
2015/2/23                     & 57077 & $3.14\pm0.03$ & $2.65\pm0.08$ \\
2015/3/26                     & 57108 & $3.09\pm0.06$ & $2.85\pm0.04$ \\
2015/4/30                      & 57143 & $3.13\pm0.07$ & $3.01\pm0.25$ \\
2015/9/24                     & 57290 & $3.12\pm0.16$ & $3.15\pm0.21$ \\
2015/10/23                    & 57319 & $3.14\pm0.10$ & $3.08\pm0.17$ \\
2015/11/30                     & 57357 & $3.14\pm0.05$ & $3.07\pm0.08$ \\
2015/12/29                    & 57385 & $3.03\pm0.06$ & $2.90\pm0.07$ \\
2016/1/13                     & 57401 & $3.12\pm0.05$ & $2.89\pm0.06$ \\
2016/2/11                     & 57430 & $2.87\pm0.07$ & $2.58\pm0.10$ \\
2016/3/1                      & 57449 & $3.02\pm0.04$ & $2.79\pm0.10$ \\
2016/4/25                     & 57504 & $3.15\pm0.13$ & $2.98\pm0.14$ \\
2016/8/23                     & 57624 & --            & $3.56\pm0.18$ \\
2016/10/18                    & 57680 & $4.24\pm0.11$ & $3.94\pm0.07$ \\
2016/11/27                    & 57720 & $3.96\pm0.24$ & $3.24\pm0.32$ \\
2016/12/28                    & 57751 & $3.42\pm0.08$ & $3.05\pm0.11$ \\
2017/3/28                     & 57841 & $3.26\pm0.05$ & $3.26\pm0.05$ \\
2017/4/20                     & 57863 & $2.92\pm0.14$ & $3.39\pm0.08$ \\
2017/5/21                     & 57895 & $3.12\pm0.09$ & $3.61\pm0.08$ \\
2017/6/17                     & 57922 & $3.68\pm0.16$ & $3.52\pm0.17$ \\
2017/10/21                    & 58048 & $3.74\pm0.03$ & $5.24\pm0.06$ \\
2017/11/4                     & 58063 & $3.72\pm0.01$ & $4.93\pm0.13$ \\
2017/11/22                    & 58080 & $3.93\pm0.13$ & $5.05\pm0.12$ \\
2018/1/13                     & 58132 & $4.35\pm0.05$ & $6.95\pm0.17$ \\
2018/2/19                     & 58169 & $4.84\pm0.10$ & $5.13\pm0.16$ \\
2018/3/2                      & 58180 & $4.11\pm0.06$ & $4.57\pm0.10$ \\
2018/3/25                     & 58203 & $4.03\pm0.07$ & $4.21\pm0.17$ \\
2018/4/19                     & 58228 & $3.75\pm0.07$ & $3.89\pm0.10$ \\
2018/5/26                     & 58265 & $3.11\pm0.10$ & $3.11\pm0.02$ \\
2018/5/31                      & 58270 & $3.26\pm0.20$ & $3.14\pm0.30$ \\
\hline
\end{tabular}
\begin{tablenotes}
\item Note. Column designation: (1) observing date in year/month/day, (2) observing date in modified Julian date (MJD), (3) flux density at 22~GHz, and (4) flux density at 43~GHz.
\end{tablenotes}
\end{threeparttable}
}
\end{table}

\subsection{Multi-frequency Data}    \label{subsec:mf data}

\subsubsection{Effelsberg 100\,m}  \label{subsubsec:Effelsberg}
The F-GAMMA program monitored about 60 blazars using various radio observatories between 2007 and 2015 \citep{Fuhrmann+2016}.
As a part of this monitoring program, the Effelsberg 100-m telescope observed CTA~102 at eight frequencies (2.64, 4.85, 8.35, 10.45, 14.6, 23.05, 32.0, and 43.0~GHz).
We made use of their flux density measurements from the period between December 2012 and December 2014 (MJD 56263--56993) in \citet{Angelakis+2019}.

\subsubsection{OVRO 40\,m}   \label{subsubsec:OVRO}
CTA~102 has been observed at 15~GHz with the Owens Valley Radio Observatory
(OVRO) 40-m radio telescope as a part of the OVRO 40-m monitoring program since 2008 \citep{Richards+2011}.
The mean cadence of these observations is about 7 days, we used the data collected from November 2012 to July 2018 (MJD 56251--58320).

\subsubsection{Mets\"{a}hovi 14\,m}
The 37~GHz observations were made with the 13.7-m diameter Mets\"{a}hovi radio telescope.
This radio telescope has been used to monitor hundreds of AGNs since the early 1980s\footnote{\url{https://www.metsahovi.fi/opendata}}.
We obtained data for the period from 2012 December to 2018 May (MJD 56266--58267).
The observations are ON--ON observations, which alternate the source and the sky in each feed horn.
A typical integration time to obtain one flux density data point is between 1200 and 1400~s.
The detection limit of our telescope at 37~GHz is on the order of 0.2~Jy under optimal conditions.
Data points with a signal-to-noise ratio < 4 are handled as non-detections.
The flux density scale was set by observations of DR 21.
Sources NGC 7027, 3C 274, and 3C 84 were used as secondary calibrators.
A detailed description of the data reduction and analysis is given in \citep{Teraesranta+1998}.
The error estimate for the flux density includes the contributions from the measurement rms and from the uncertainty in the absolute calibration.

\subsubsection{IRAM 30\,m}
The measurements from the IRAM 30-m Telescope that we present in this paper were obtained as part of the POLAMI (Polarimetric Monitoring of AGN at Millimetre Wavelengths) Program\footnote{\url{https://polami.iaa.es}} \citep{Agudo+2018a, Agudo+2018b, Thum+2018}.
The POLAMI program is a long-term observation project to monitor the full polarimetric of a sample of bright millimeter AGNs including CTA 102 at 86 and 230~GHz frequencies.
This program has been operating since the end of 2006, and since then it has maintained a monitoring time sampling period between $\sim$2 weeks and 1 month depending on the monitored source.
The setup employed in these observations has been detailed in \citet{Thum+2008}, while details about the data reduction and calibration used have been presented in \citet{Agudo+2010, Agudo+2014, Agudo+2018b}.

\subsubsection{SMA} \label{subsubsec:SMA}
The 230~GHz (1.3~mm) flux density data were obtained by the Submillimeter Array (SMA) near the summit of Mauna Kea (Hawaii).
The available data spans the time period from 2013 February to 2018 April (MJD 56328--58233).
CTA~102 is included in an ongoing monitoring program at the SMA to determine the fluxes of compact extragalactic radio sources that can be used as mm-wavelength calibrators \citep{Gurwell+2007}.
The available potential calibrators are observed from time to time for 3 to 5~minutes, and the measured source signal strength is calibrated against known standards, typically solar system objects (Titan, Uranus, Neptune, or Callisto).
Data from this program are updated regularly and are available at the SMA website \footnote{\url{https://sma1.sma.hawaii.edu/callist/callist.html}}.

\subsubsection{ALMA}    \label{subsubsec:ALMA}
We made use of the Atacama Large millimeter/submillimeter Array (ALMA) data from the ALMA Calibrator Source Catalogue website\footnote{\url{https://almascience.eso.org/alma-data/calibrator-catalogue}}.
CTA~102 was observed at Band 3 (91.5, 95.7, 97.5, 98.2, 99.4, 103.5, and 109.7~GHz), Band 6 (233~GHz), and Band 7 (337.5, 343.2, and 343.5~GHz).
The available data spans a time period from 2013 June to 2017 September (MJD 56454--58025).

\subsubsection{VLBA}   \label{subsubsec:VLBA}
CTA 102 was observed as a part of the Monitoring of Jets in Active Galactic Nuclei with VLBA Experiments (MOJAVE) program.
This program is a long-term observational program to study the structure and evolution of relativistic outflows in AGNs using the very long baseline array (VLBA) at 15~GHz \citep{Lister+2005}.
The data are publicly available from the program website\footnote{\url{http://www.physics.purdue.edu/MOJAVE/}}.
We used the calibrated data for CTA~102 \citep[see,][for calibration details]{Lister+2009a} in order to obtain information about the core components (e.g., total flux density, size, etc.) from 2012 December--2013 July (MJD 56271--56333).
We parameterized the maps using circular Gaussians fitted directly to the \emph{uv}-data using the model-fit task in the interferometric imaging software DIFMAP \citep{Shepherd+1997}.
The uncertainties in the Gaussian components were estimated following the method in \citet{Lister+2016}.

The source was also observed as a part of the VLBA-BU-BLAZAR monitoring program.
This program observes many $\gamma$-ray bright blazars approximately monthly using the VLBA at 43~GHz \citep{Jorstad+2005, Jorstad+2017}.
The data are publicly available from the program website\footnote{\url{https://www.bu.edu/blazars/VLBAproject.html}}.
We used the calibrated data for the source \citep[see,][for calibration details]{Jorstad+2017} in order to obtain information about the core components (e.g., total flux density, size, etc.) from 2013 April--2018 April (MJD 56398--58227).
In order to do this, we performed our analysis in the same way as for the MOJAVE data.
Following \citep{Jorstad+2017}, we estimated the uncertainties in the Gaussian components.

\begin{figure*}
    \centering
    \includegraphics[width=1.0\textwidth]{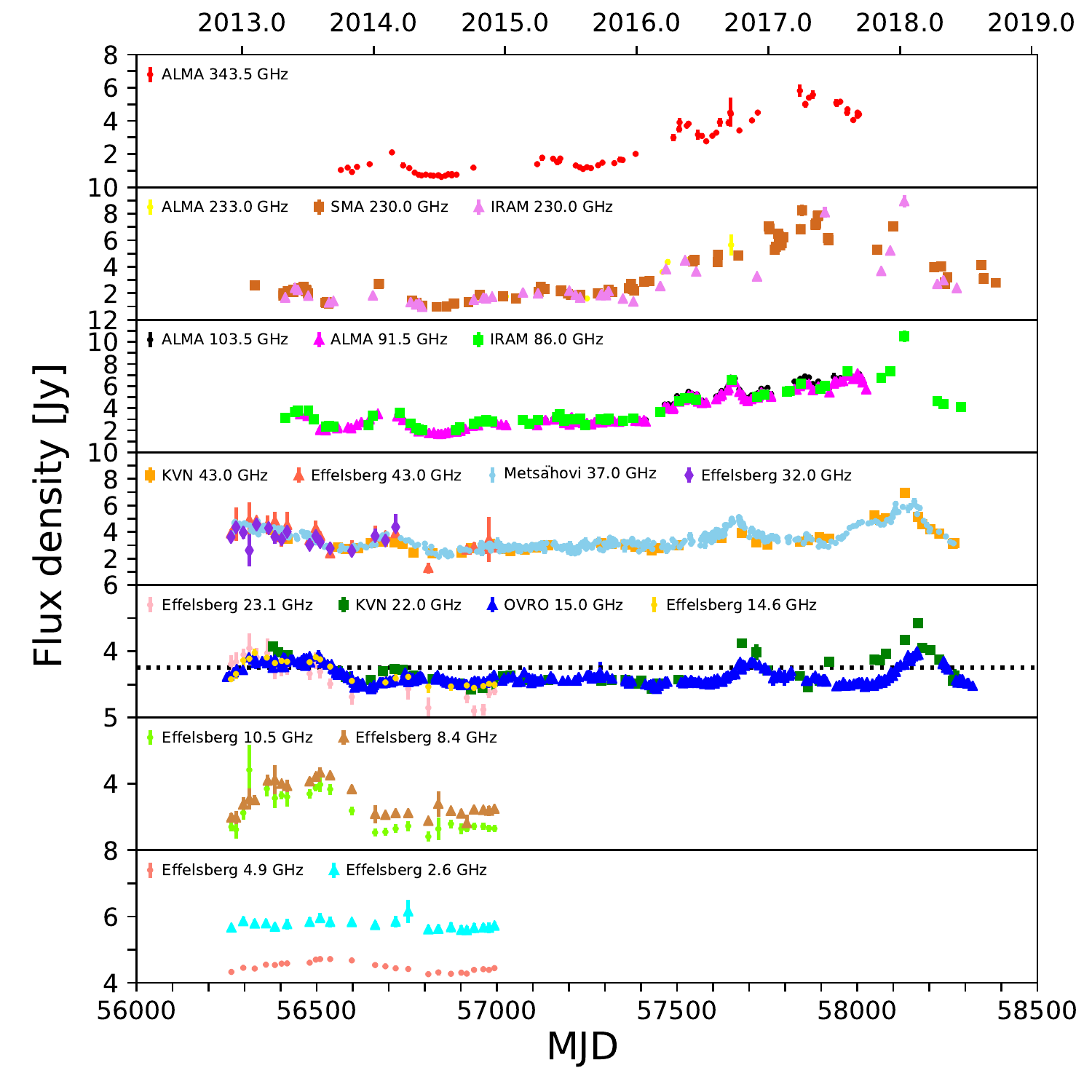}
    \caption{Light curves from CTA~102 observed between 2012 November and 2018 September (MJD 56251 to 58384) at 2.6--343.5~GHz. The light curves at different frequencies are shown with different colors and symbols (see the label). The black dotted line for the 15.0~GHz light curve indicates the flux density threshold (described in the text).}
    \label{fig:light curves}
\end{figure*}

\section{Results} \label{sec:results}

\subsection{Radio Light Curves}
\label{sec: Radio LCs}
Figure \ref{fig:light curves} shows the multi-frequency light curves from CTA~102 at the radio frequencies of 2.6--343.5~GHz.
The light curves cover a $\sim$5.5-year time span, from 2012 November to 2018 September (MJD 56251--58384).
The OVRO 15~GHz light curve is well sampled with a mean cadence of $t_{\rm int}$ $\sim$ 7 days, as seen in Table \ref{table: all observations}.
This well-sampled light curve enables us to clearly see an active state period, in which the source becomes brighter and more variable than during other periods.
We considered a threshold using the median plus the standard deviation of the flux density for defining the active state \citep[e.g.,][]{Algaba+2018a}.
We found three periods that counted as active states in the 15~GHz light curve: 2013 January 20--2013 September 6 (MJD 56312--56541), 2016 October 9--2016 December 5 (MJD 57670--57728), and 2017 December 23--2018 May 4 (MJD 58110--58243).
During these periods, the light curve peaks on 2013 July 31 (MJD 56504), 2016 November 16 (MJD 57708), and 2018 February 17 (MJD 58166) where respective flux densities of 3.88, 3.69, and 3.99~Jy, were reached.

We found that such active states are also reflected in the light curves at other frequencies. 
For example, the first active state was found in the light curves at $\sim$8--43~GHz, the second and third active states were found in the light curves at $>$ 15~GHz.
In the light curves at $>$ 86~GHz, the flux density of the source begins to gradually increase in the middle of 2015, yielding a mean flux density that has increased by a factor of 1.8 to 3.3.
In the following section, Section \ref{subsec:decomposition}, we used the flux variability seen in the 15~GHz light curve to identify flares.

\subsection{Decomposition of Radio Flares}  \label{subsec:decomposition}
The variations in the flux densities of blazars have been attributed to several flares that occurred over the time-scale of days to months \citep{Valtaoja+1999, Hovatta+2009, Liodakis+2017}.
Of these observed flares, most have been found to overlap somewhat, making it difficult to directly identify individual flares directly.
In order to decompose the flares, we employ the Markov Chain Monte Carlo (MCMC) technique to identify individual flares and to constrain the best-fitting model parameters for the flares.

\begin{figure}
    \centering
    \includegraphics[width=0.45\textwidth]{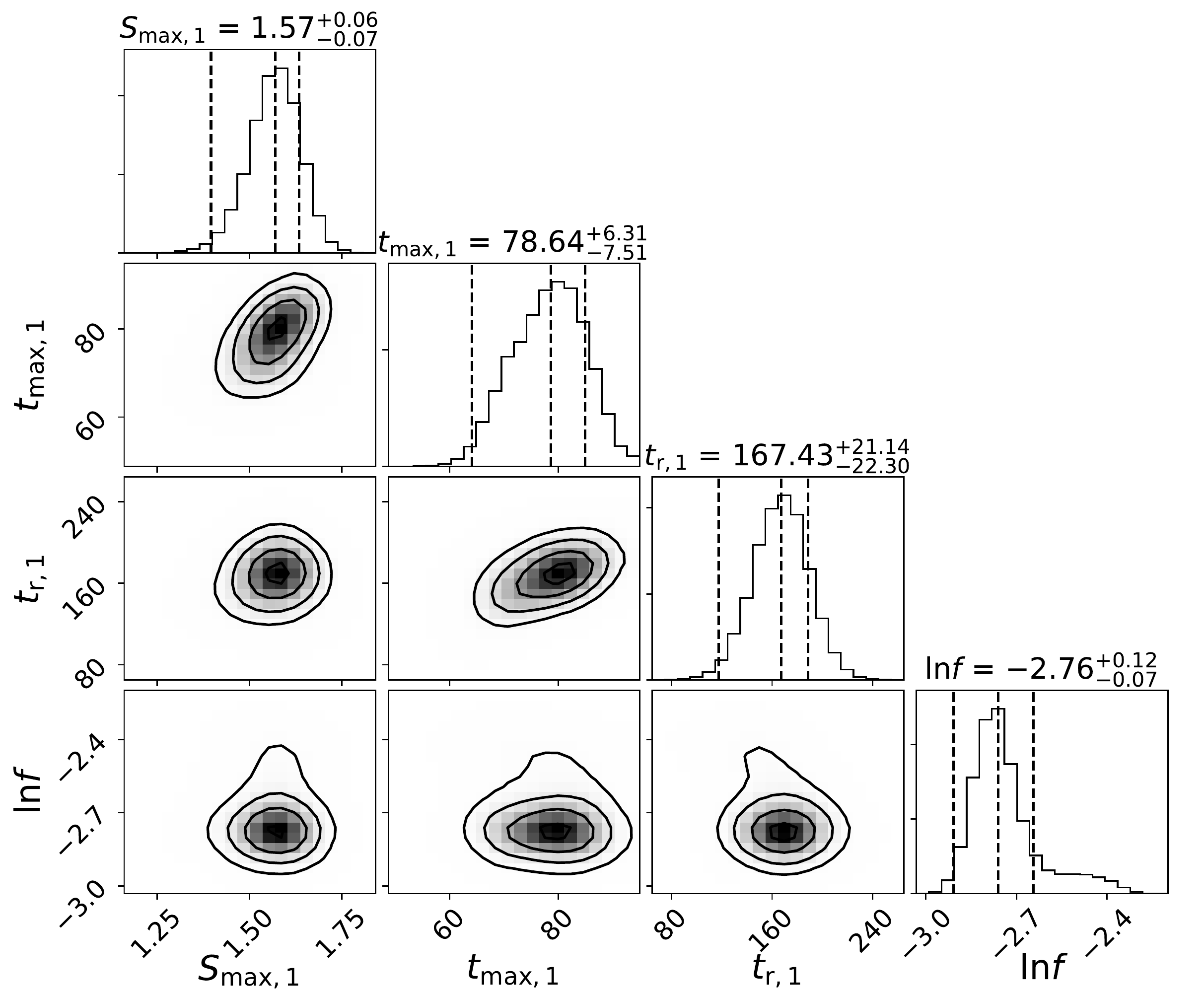}
    \caption{Posterior distribution of fitting parameters. The fitting parameters are the peak flux density of the flare $S_{\mathrm{max},i}$, the time of the peak flux density of the flare $t_{\mathrm{max,i}}$, the rise time-scale of the flare $t_{\mathrm{r,i}}$ ($i$ denotes the flare number), and the scaling factor $f$ that is described in the text. The dashed lines are the 16th, 50th, and 84th quantiles of the Gaussian distribution from each fitting parameter. Note that this plot is an example using the first of the seven fitted flares.}
    \label{fig:corner plot}
\end{figure}

\subsubsection{Analysis}    \label{subsubsec: decomp analysis}
One of the requirements for a detailed flare decomposition analysis is a short cadence in the light curves, specifically in the well-sampled light curves.
Among the multi-frequency light curves, we found that the OVRO and Mets\"{a}hovi light curves had mean cadences of $t_{\rm int}$ $\sim$ 7~days and 5~days, respectively, making these the most suitable ones for this analysis.
In this paper, we simply chose to use the OVRO light curve for our decomposition analysis although the comparison of the light curves itself is still of significant interest (and will be presented in an upcoming paper).
A flare is characterized by three parameters -- its maximum amplitude, the time of the flare maximum, and the rise time-scale.
The ratio of the decay to the rise time-scale is fixed at 1.3 \citep[, following][]{Valtaoja+1999, Hovatta+2009, Kravchenko+2020}.
An exponential curve model was used following \citet{Valtaoja+1999}:
\begin{equation}	\label{eq: radio flares}
    S(t) = \begin{cases}
    S_\mathrm{max}e^{(t - t_\mathrm{max})/t_\mathrm{r}}, & \text{$t < t_\mathrm{max}$}, \\
    S_\mathrm{max}e^{(t_\mathrm{max} - t)/1.3t_\mathrm{r}}, & \text{$t > t_\mathrm{max}$},
    \end{cases}
\end{equation}
where $S_\mathrm{max}$ is the maximum amplitude of a flare in Jy, $t_\mathrm{max}$ is the time of the maximum amplitude of the flare in days, and $t_\mathrm{r}$ is the rise time-scale of the flare in days.
Before performing the decomposition analysis, a constant quiescent flux density was subtracted from all the flux densities of the light curve \citep[e.g.,][]{Valtaoja+1999}.
The quiescent flux density was estimated to be 2.17~Jy using the quiescent spectrum (see Section \ref{subsec: source spectra}, for more details).

In order to fit the exponential model to the light curve, an initial number of flares must be determined first.
We tried to determine the number of flares in the following manner.
The initial model ($S_{\rm init}$, $t_{\rm init}$, and $t_{\rm r, init}$) of the first flare was determined based on the first peak ($S_{\rm peak}$ and $t_{\rm peak}$) in the light curve.
The peak flux density and time of the initial model were determined as $S_{\rm init}=S_{\rm peak}$ and $t_{\rm init}=t_{\rm peak}$.
The rise time-scale of the initial model $t_{\rm r, init}$ was simply determined by observing when the residual gets lower than the three sigma level ($3 \times \sigma_\mathrm{S}$) defined as $\sigma_{\rm S} = \sqrt{\sum_{i=1}^N (\sigma_{\mathrm{stat},i}^2 + \sigma_{\mathrm{sys}, i}^2)/N}$, where $\sigma_{\rm stat}$ is the statistical flux uncertainty, $\sigma_{\rm sys}$ is the 5~{per cent} systematic flux uncertainty, and $N$ denotes the number of the data.
The determined initial model for the first flare was subtracted from the light curve.
An initial model of the second flare was determined based on the first peak in the remaining data following the same procedure that was used for the first flare.
The determined initial model of the second flare was also subtracted.
We repeated this procedure for the remaining data until flux densities higher than the three-sigma level were not seen.
Finally, we determined the initial values for the 21 fitting parameters that compose the seven flares.

\begin{figure*}
    \centering
    \includegraphics[width=1.0\textwidth]{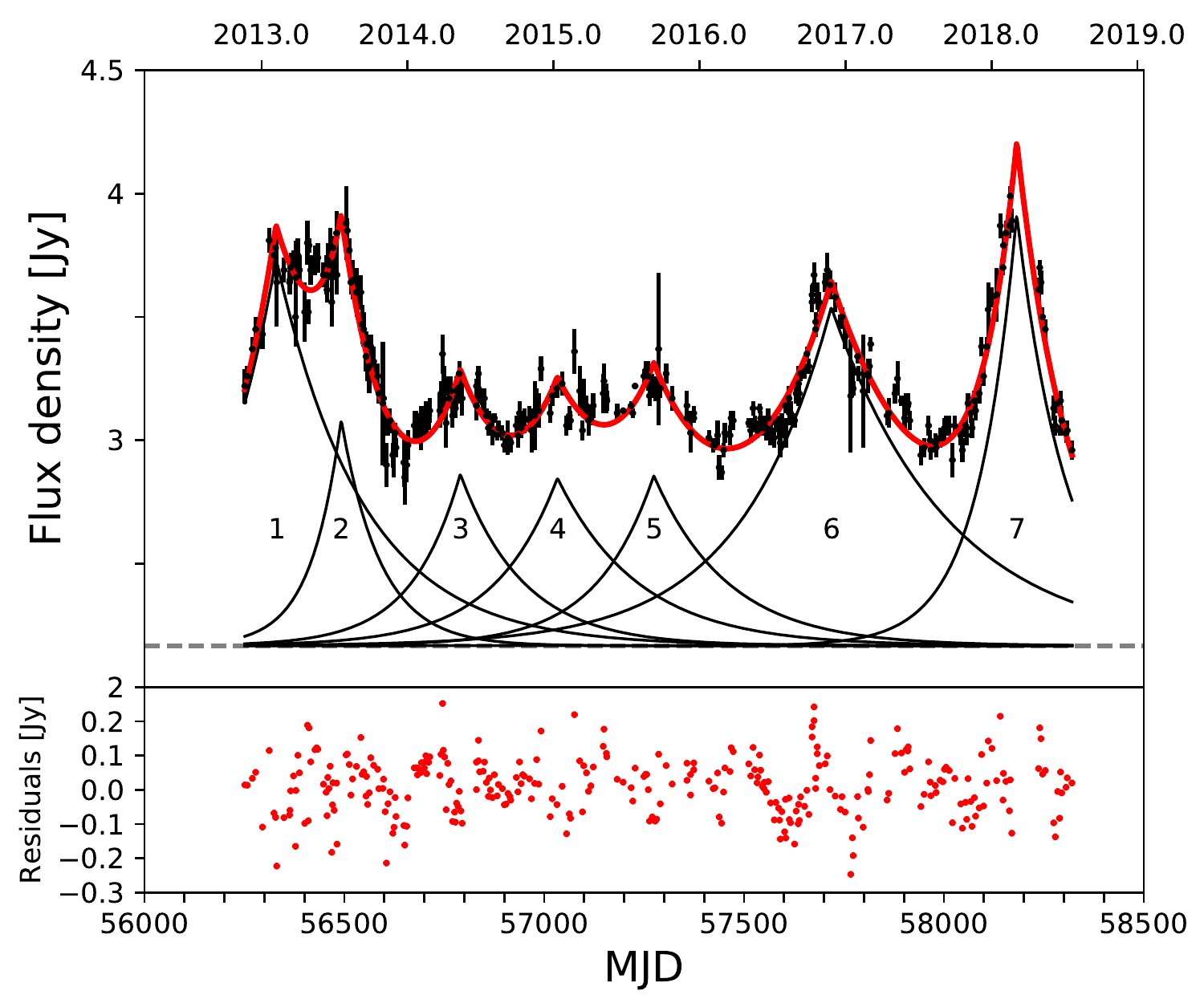}
    \caption{The upper panel shows the light curve from CTA~102 at 15~GHz (black points) decomposed into seven exponential flares (thick red solid line - the sum of the seven flares, thin black solid line - seven flares individually) at 15~GHz. The gray dashed line indicates the quiescent flux density of 2.17~Jy. The designation of the flares is also shown in the figure. In the lower panel, the red circles are the residuals between the data and the flare models.}
    \label{fig:decomp result}
\end{figure*}

The model flares with the initial parameters (Eq. \ref{eq: radio flares}) were fitted to the light curve using the MCMC technique.
A Python package named {\emph{emcee}} \citep{Foreman-Mackey+2013} was used to sample the posterior distributions of the three parameters for each flare: $S_\mathrm{max}$, $t_\mathrm{max}$, and $t_\mathrm{r}$.
We included an additional fitting parameter $f$ to correct for the underestimated variance in the likelihood function\footnote{\url{https://emcee.readthedocs.io/en/stable/}}.
The ranges of uniform priors were set to as (1) $3 \times \sigma_{\rm S} < S_\mathrm{max} < S_\mathrm{max,init} + 3 \times \sigma_{\rm S}$, (2) $t_\mathrm{max} < t_\mathrm{max,init} + 5 \times t_{\rm int}$, (3) $t_{\rm r} < t_\mathrm{r,init} + 200$~days, and (4) $-10 < \ln{f} < 1$, where $S_\mathrm{max,init}$, $t_\mathrm{max,init}$, and $t_\mathrm{r,init}$ are the initial values for the fitting parameters $S_\mathrm{max}$, $t_\mathrm{max}$, and $t_\mathrm{r}$, respectively.
A Gaussian distribution of each fitting parameter was characterized by obtaining 600,000 MCMC samples.
The uncertainties in the parameters are determined based on the 68~{per cent} confidence intervals in the distributions of the parameters.
Figure \ref{fig:corner plot} shows a corner plot as an example of the distributions of the best-fitting parameters.
The results of the decomposition analysis are summarized in Table~\ref{table: flare parameters}.
In Figure \ref{fig:decomp result}, the observed data and the best-fitting model are presented.

\subsubsection{Variability Parameters Estimation}   \label{subsec:variability parameter}
We quantify the variability characteristics of the source such as the variability time-scales, variability brightness temperatures, and the sizes of the emission region.
The variability time-scales are defined by the rise time of a flare ($t_\mathrm{r}$) obtained from the decomposition analysis, as shown in \citet[][]{Hovatta+2009} and \citet{Liodakis+2017}.
Note that the period of the Flare~1 is out of the time range of the data considering its variability time-scale.
This may lead to changes in the variability parameters of Flare~1.

\begin{table*}
\caption{Physical parameters of radio flares from CTA 102}
\label{table: flare parameters}
\centering
{
\begin{threeparttable}
\begin{tabular}{ccccccc}
\hline
Flare\_num & $t_\mathrm{max}$ & $t_\mathrm{max}$ (MJD) & $S_\mathrm{max}$ (Jy)  & $t_\mathrm{r}$ (days) & $T_\mathrm{b,var}$ ($10^{13}$ K) & $d_\mathrm{var}$ (mas) \\
(1) & (2) & (3) & (4) & (5) & (6) & (7) \\
\hline
1  & $2013/2/6^{+6}_{-7}$   & $56329^{+6}_{-7}$  & $1.57^{+0.06}_{-0.07}$ & $167^{+21}_{-22}$ & $1.03^{+0.26}_{-0.28}$ & $0.29\pm 0.06$         \\
2  & $2013/7/19^{+3}_{-4}$  & $56492^{+3}_{-4}$  & $0.92^{+0.10}_{-0.08}$ & $76^{+18}_{-14}$  & $2.90^{+1.41}_{-1.10}$ & $0.13^{+0.04}_{-0.03}$ \\
3  & $2014/5/13\pm 6$       & $56790\pm 6$       & $0.70^{+0.07}_{-0.04}$ & $122^{+25}_{-18}$ & $0.86^{+0.36}_{-0.26}$ & $0.21^{+0.06}_{-0.05}$ \\
4  & $2015/1/11^{+12}_{-9}$ & $57033^{+12}_{-9}$ & $0.68^{+0.05}_{-0.03}$ & $158^{+25}_{-26}$ & $0.50^{+0.16}_{-0.17}$ & $0.28\pm 0.07$         \\
5  & $2015/9/9^{+13}_{-8}$  & $57274^{+13}_{-8}$ & $0.69^{+0.06}_{-0.03}$ & $143^{+21}_{-23}$ & $0.62^{+0.19}_{-0.20}$ & $0.25 \pm 0.06$         \\
6  & $2016/11/26^{+3}_{-4}$ & $57718^{+3}_{-4}$  & $1.37\pm 0.03$         & $227^{+17}_{-18}$ & $0.49^{+0.07}_{-0.08}$ & $0.40 \pm 0.08$         \\
7  & $2018/3/5\pm 2$        & $58182\pm 2$       & $1.75\pm 0.07$         & $98^{+11}_{-7}$   & $3.33^{+0.76}_{-0.50}$ & $0.17^{+0.04}_{-0.03}$         \\
\hline
\end{tabular}
\begin{tablenotes}
\item Note. Column designation: (1) flare number, (2) time of the maximum amplitude of the flare in year/month/day, (3) time of the maximum amplitude of the flare in modified Julian date, (3) maximum amplitude of the flare in Jy, (4) rising time-scale in days, (6) brightness temperature in $10^{13}$~K, and (7) emitting region size in mas.
\end{tablenotes}
\end{threeparttable}
}
\end{table*}

We estimate the variability brightness temperature ($T_\mathrm{b, var}$):
\begin{equation}
    T_\mathrm{b, var} = 1.47 \times 10^{13}\frac{{D_\mathrm{L}^2} S_\mathrm{max}}{{\nu}^2{t_\mathrm{r}^2}{(1+z)}^4}~[K],
\end{equation}
where $D_\mathrm{L}$ is the luminosity distance in Mpc, $S_\mathrm{max}$ is defined in Eq. \ref{eq: radio flares}, $\nu$ is the observing frequency in GHz, and $z$ is the redshift of the source.
We obtained $T_\mathrm{b, var}$ for the seven flares to be in the range of $\sim$(0.5--3.0) $\times 10^{13}$~K and provide exact numbers in Table \ref{table: flare parameters}.

The rapid variability time-scale imply highly compact emission regions if these variations are intrinsic properties of the source.
Assuming the source has a spherical brightness distribution with isotropic expansion, the light travel time argument implies a diameter $d \leq 2 \times c \Delta t$ of the emitting region, where $\Delta t$ is the time interval of the expansion.
Then, we can obtain the emitting region size ($d_\mathrm{var}$):
\begin{equation}
    d_\mathrm{var} = 0.35{\frac{t_\mathrm{r}}{D_\mathrm{L}}}\delta(1+z)~[\mathrm{mas}].
\end{equation}

We take a Doppler factor as $\delta = 17 \pm 3$, which is the variability Doppler factor of a moving jet component close to the VLBI core at 15~GHz \citep{Fromm+2013a}.
The value of $d_\mathrm{var}$ is estimated to be in a range of 0.13--0.40~mas.
The mean and the standard deviation values are 0.25 and 0.08~mas, respectively.
The emitting region size $d_\mathrm{var}$ is broadly consistent with the VLBI core size.
We will investigate this in Section \ref{subsec: compact region}.
The results of the variability parameters are summarized in Table \ref{table: flare parameters}.

\subsection{Source Spectra} \label{subsec: source spectra}

The single-dish observations may include emissions from extended components which could be resolved out by the VLBI observations.
Hence, one should be careful when using the flux densities from single-dish and VLBI observations together for spectral analysis.
We compared the single-dish flux density and the VLBI total flux density at the same observation frequency in nearby epochs separated by less than 10~days to confirm whether it is possible for those flux densities to be used together.

\begin{figure*}
    \centering
    \includegraphics[width=\textwidth]{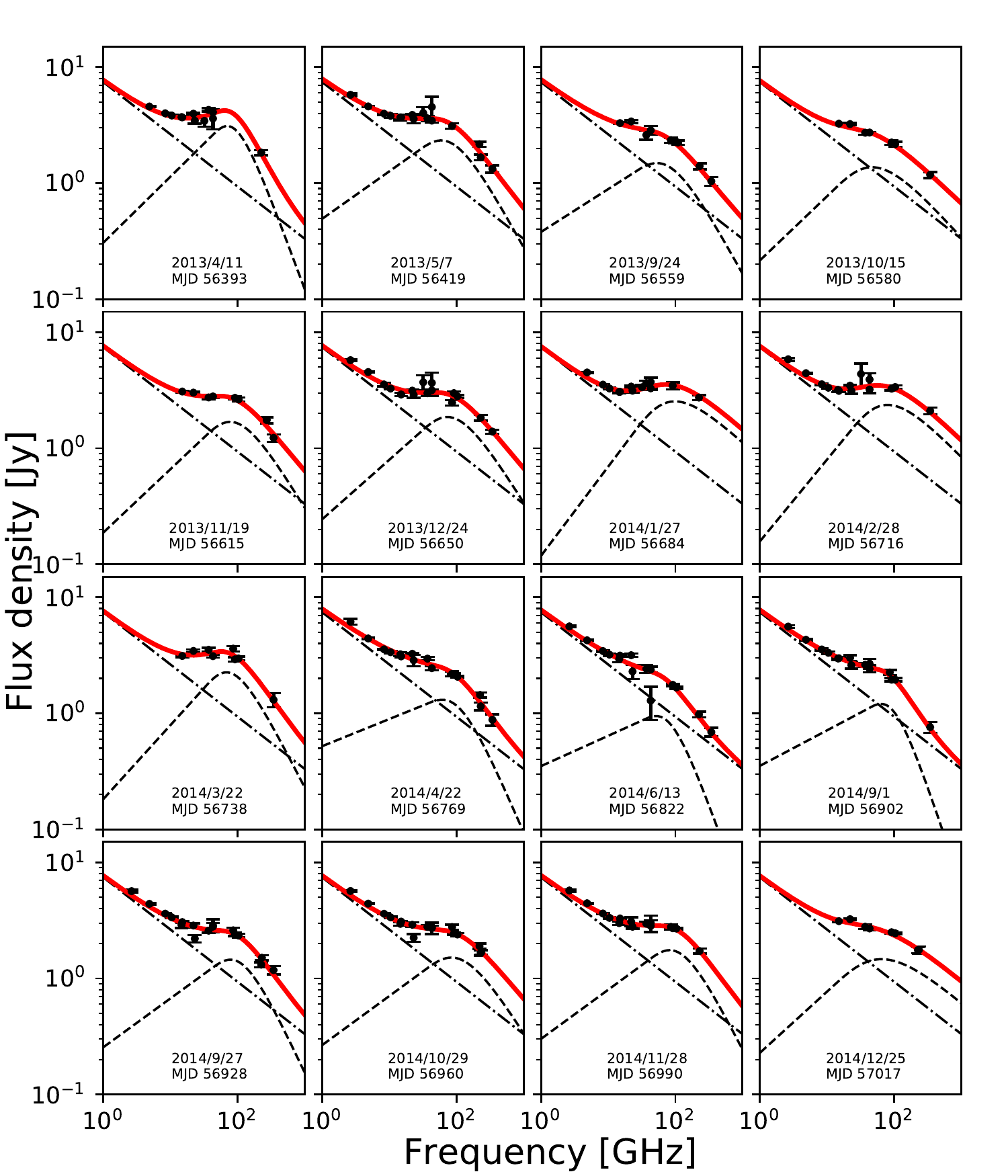}
    \caption{Spectral results for the SSA region. A total of 33 epochs out of 47 epochs were selected (see text for details). The spectral results come from between 2013 April 11 (MJD 56393) and 2018 April 19 (MJD 58228). In each panel, the observation date in UT and the corresponding time in MJD are labeled in the bottom center of each graph. The black points indicate the observed data with errors. The quiescent spectrum (thin black dot-dashed line), the flaring spectrum (thin black dashed line), and the total spectrum (thick red solid line) are also plotted.}
    \label{fig:SSA spectra}
\end{figure*}

\begin{figure*}\ContinuedFloat
    \centering
    \includegraphics[width=\textwidth]{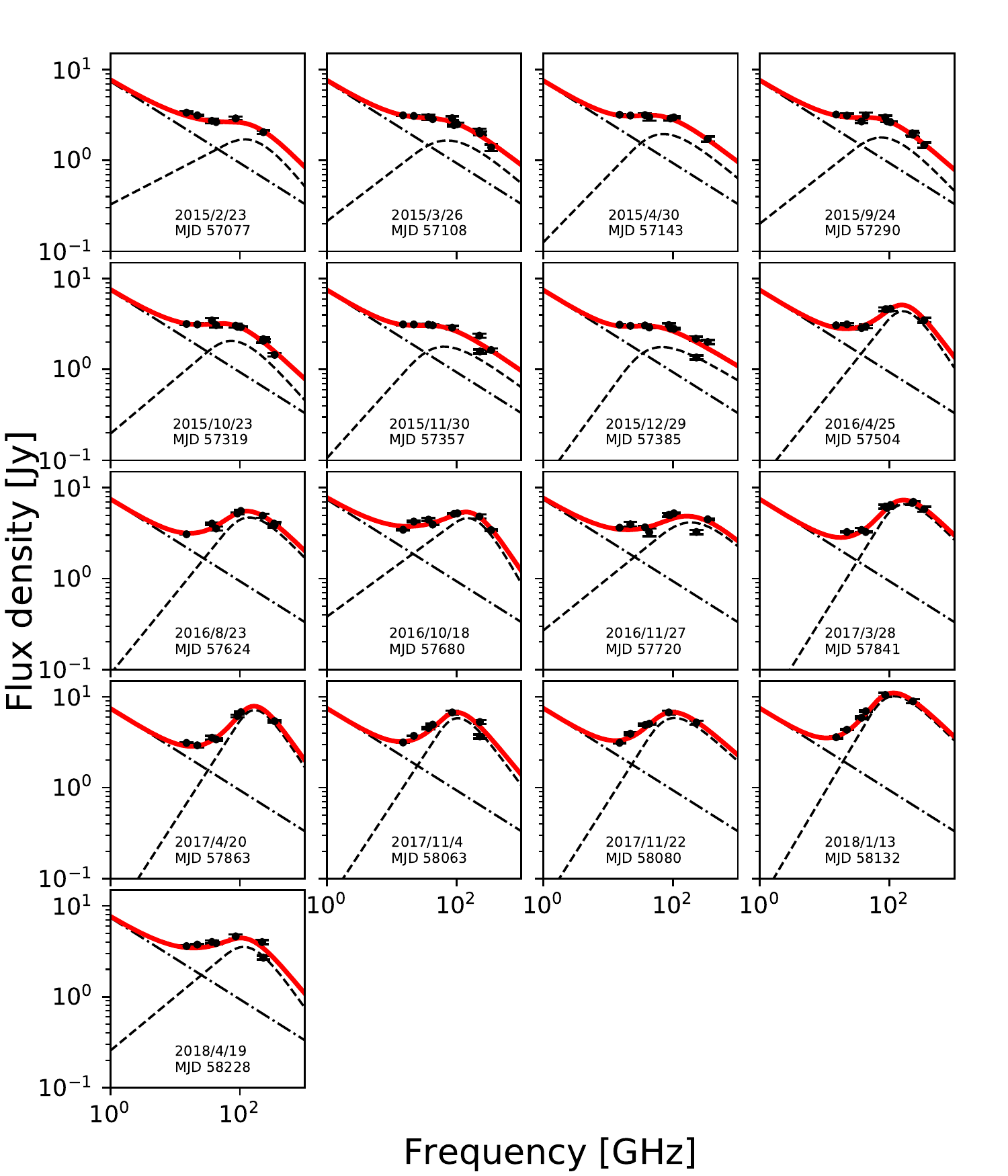}
    \caption{continued.}
\end{figure*}

\begin{table*}
\caption{Results of the SSA spectrum fitting analysis}
\label{table: SA - turnover parameters}
\centering
{
\begin{threeparttable}
\begin{tabular}{cccccc}
\hline
Date & MJD & $\alpha_\mathrm{t}$ & $S_\mathrm{m}$~(Jy) & $\nu_\mathrm{m}$~(GHz) & $\alpha_\mathrm{0}$ \\
(1) & (2) & (3) & (4) & (5) & (6) \\
\hline
2013/4/11  & 56393 & $0.57^{+0.15}_{-0.12}$ & $3.07^{+0.58}_{-0.49}$  & $62.94^{+14.21}_{-16.35}$  & $-1.56^{+0.62}_{-0.63}$ \\
2013/5/7   & 56419 & $0.41^{+0.20}_{-0.13}$ & $2.31^{+0.23}_{-0.20}$  & $50.92^{+16.57}_{-10.78}$  & $-1.06^{+0.32}_{-0.49}$ \\
2013/9/24  & 56559 & $0.37^{+0.71}_{-0.26}$ & $1.47^{+0.21}_{-0.17}$  & $45.00^{+19.34}_{-12.75}$  & $-1.06^{+0.49}_{-0.72}$ \\
2013/10/15 & 56580 & $0.55^{+0.93}_{-0.42}$ & $1.37^{+0.19}_{-0.16}$  & $42.82^{+20.66}_{-12.94}$  & $-0.65^{+0.32}_{-0.66}$ \\
2013/11/19 & 56615 & $0.55^{+0.47}_{-0.28}$ & $1.68^{+0.20}_{-0.20}$  & $69.86^{+10.87}_{-14.68}$  & $-0.96^{+0.39}_{-0.41}$ \\
2013/12/24 & 56650 & $0.51^{+0.25}_{-0.17}$ & $1.85^{+0.16}_{-0.15}$  & $65.35^{+11.55}_{-9.52}$   & $-0.94^{+0.26}_{-0.32}$ \\
2014/1/27  & 56684 & $0.77^{+0.41}_{-0.23}$ & $2.53^{+0.39}_{-0.26}$  & $94.25^{+23.08}_{-16.75}$  & $-0.54^{+0.31}_{-0.68}$ \\
2014/2/28  & 56716 & $0.71^{+0.31}_{-0.19}$ & $2.36^{+0.19}_{-0.19}$  & $75.34^{+7.54}_{-10.72}$   & $-0.60^{+0.21}_{-0.23}$ \\
2014/3/22  & 56738 & $0.64^{+0.49}_{-0.29}$ & $2.24^{+0.26}_{-0.23}$  & $62.27^{+12.22}_{-10.82}$  & $-1.13^{+0.43}_{-0.49}$ \\
2014/4/22  & 56769 & $0.24^{+0.25}_{-0.13}$ & $1.28^{+0.16}_{-0.14}$  & $47.94^{+19.14}_{-12.80}$  & $-1.36^{+0.49}_{-0.68}$ \\
2014/6/13  & 56822 & $0.26^{+0.32}_{-0.16}$ & $0.93^{+0.18}_{-0.15}$  & $41.72^{+17.17}_{-13.19}$  & $-1.61^{+0.67}_{-0.64}$ \\
2014/9/1   & 56902 & $0.31^{+0.22}_{-0.15}$ & $1.18^{+0.17}_{-0.14}$  & $52.73^{+14.61}_{-11.41}$  & $-1.73^{+0.58}_{-0.55}$ \\
2014/9/27  & 56928 & $0.42^{+0.21}_{-0.15}$ & $1.44^{+0.15}_{-0.15}$  & $67.51^{+11.46}_{-11.39}$  & $-1.24^{+0.35}_{-0.40}$ \\
2014/10/29 & 56960 & $0.42^{+0.22}_{-0.15}$ & $1.50^{+0.14}_{-0.13}$  & $73.01^{+8.86}_{-11.04}$   & $-0.93^{+0.36}_{-0.38}$ \\
2014/11/28 & 56990 & $0.42^{+0.19}_{-0.13}$ & $1.74^{+0.15}_{-0.14}$  & $71.33^{+9.97}_{-11.28}$   & $-1.15^{+0.43}_{-0.49}$ \\
2014/12/25 & 57017 & $0.53^{+0.95}_{-0.37}$ & $1.46^{+0.16}_{-0.14}$  & $59.67^{+16.52}_{-17.74}$  & $-0.51^{+0.35}_{-0.69}$ \\
2015/2/23  & 57077 & $0.37^{+0.36}_{-0.22}$ & $1.69^{+0.33}_{-0.23}$  & $100.44^{+26.85}_{-23.76}$ & $-0.95^{+0.74}_{-1.00}$ \\
2015/3/26  & 57108 & $0.55^{+0.76}_{-0.34}$ & $1.66^{+0.16}_{-0.15}$  & $65.61^{+13.59}_{-15.72}$  & $-0.63^{+0.33}_{-0.50}$ \\
2015/4/30   & 57143 & $0.74^{+0.69}_{-0.39}$ & $1.95^{+0.21}_{-0.21}$  & $67.67^{+12.11}_{-13.84}$  & $-0.61^{+0.30}_{-0.37}$ \\
2015/9/24  & 57290 & $0.56^{+0.52}_{-0.29}$ & $1.78^{+0.15}_{-0.15}$  & $67.63^{+11.81}_{-13.12}$  & $-0.77^{+0.31}_{-0.37}$ \\
2015/10/23 & 57319 & $0.60^{+0.51}_{-0.28}$ & $2.06^{+0.16}_{-0.15}$  & $67.63^{+11.28}_{-11.81}$  & $-0.83^{+0.29}_{-0.37}$ \\
2015/11/30  & 57357 & $0.80^{+0.81}_{-0.46}$ & $1.78^{+0.30}_{-0.26}$  & $62.15^{+15.37}_{-16.50}$  & $-0.53^{+0.29}_{-0.44}$ \\
2015/12/29 & 57385 & $0.97^{+0.81}_{-0.58}$ & $1.77^{+0.26}_{-0.22}$  & $66.20^{+12.96}_{-16.09}$  & $-0.43^{+0.25}_{-0.34}$ \\
2016/4/25  & 57504 & $0.90^{+0.22}_{-0.16}$ & $4.37^{+0.76}_{-0.57}$  & $152.84^{+21.33}_{-18.47}$ & $-1.13^{+0.50}_{-0.72}$ \\
2016/8/23  & 57624 & $0.88^{+0.32}_{-0.18}$ & $4.71^{+0.43}_{-0.34}$  & $133.28^{+14.55}_{-13.83}$ & $-0.77^{+0.32}_{-0.43}$ \\
2016/10/18 & 57680 & $0.52^{+0.16}_{-0.10}$ & $4.58^{+0.40}_{-0.42}$  & $132.19^{+20.42}_{-18.26}$ & $-1.31^{+0.62}_{-0.58}$ \\
2016/11/27 & 57720 & $0.58^{+0.30}_{-0.19}$ & $4.14^{+0.71}_{-0.54}$  & $172.33^{+68.04}_{-41.70}$ & $-0.68^{+0.43}_{-0.85}$ \\
2017/3/28  & 57841 & $1.15^{+0.19}_{-0.15}$ & $6.60^{+0.45}_{-0.36}$  & $167.45^{+17.76}_{-10.19}$ & $-0.75^{+0.29}_{-0.34}$ \\
2017/4/20  & 57863 & $1.11^{+0.19}_{-0.14}$ & $7.15^{+1.13}_{-1.05}$  & $162.69^{+18.83}_{-13.00}$ & $-1.13^{+0.55}_{-0.60}$ \\
2017/11/4  & 58063 & $1.09^{+0.45}_{-0.26}$ & $5.82^{+1.38}_{-1.01}$  & $101.13^{+18.97}_{-16.43}$ & $-0.99^{+0.57}_{-0.83}$ \\
2017/11/22 & 58080 & $1.13^{+0.42}_{-0.28}$ & $5.85^{+1.05}_{-0.68}$  & $103.43^{+17.96}_{-13.43}$ & $-0.66^{+0.36}_{-0.71}$ \\
2018/1/13  & 58132 & $1.21^{+0.24}_{-0.20}$ & $10.18^{+1.58}_{-1.02}$ & $115.45^{+15.05}_{-10.81}$ & $-0.71^{+0.29}_{-0.58}$ \\
2018/4/19  & 58228 & $0.59^{+0.38}_{-0.22}$ & $3.52^{+0.71}_{-0.54}$  & $103.97^{+21.13}_{-21.21}$ & $-1.05^{+0.73}_{-0.99}$ \\
\hline
\end{tabular}
\begin{tablenotes}
\item Note. Column designation: (1) observing date in year/month/day, (2) observing date in modified Julian date, (3) optically thick spectral index, (4) peak flux density in Jy, (5) turnover frequency in GHz, and (6) optically thin spectral index
\end{tablenotes}
\end{threeparttable}
}
\end{table*}

The 15~GHz OVRO data were used to obtain the single-dish flux density.
For the VLBI total flux density, the 15~GHz VLBA data from the MOJAVE program were used.
In 3 epochs, the single-dish flux density and the VLBI total flux density showed flux density ratios of $\sim$11--21~{per cent}, which is much larger than typical OVRO errors of $\sim$1~{per cent} and VLBA errors of $\sim$5~{per cent}.
This implies that the single-dish flux density may include emissions from extended components which could be resolved out by the VLBI observations.
Thus, we cannot make use of the single-dish flux density together with the VLBI flux density for a multi-frequency spectral analysis.
We compared the single-dish flux density and the VLBI total flux density at 43~GHz.
In order to do this, we used the KVN single-dish and VLBA data.
Over 18 epochs, the ratio of the single-dish flux density to the VLBI total flux density was found to be $\sim$2--38~{per cent} with a mean of $\sim$24~{per cent}.
This supports the existence of a difference between the single-dish flux density and the VLBI total flux density.

We also compared the single-dish flux density and the array flux density.
We used the IRAM (single-dish), ALMA (array), and SMA (array) data at 230~GHz for the comparison.
The flux density difference is less than 7~{per cent}, which is comparable to typical errors of $\sim$6~{per cent}.
This indicates that the single-dish and array flux densities are comparable to each other and that the result of the spectral analysis is not sensitive to combining the single-dish and array flux densities.
Therefore, we used the combination of single-dish and array flux densities in our results.

To perform spectral analysis, we first concentrate on epochs for which the KVN single-dish data are available.
We search for other multi-frequency data from the closest epochs within 15~days of the KVN single-dish data and then consider those data to be quasi-simultaneous.
The flux densities from the other multi-frequency data were analyzed together with the flux density of the KVN single-dish data.
We restricted our analysis to epochs where at least five data points across different frequencies could be obtained.
We investigated the spectral properties of the synchrotron-self absorption (SSA) region using the SSA model detailed in \citep{Turler+2000, Fromm+2011, Rani+2013}:
\begin{equation}    \label{eq:SSA}
    S_{\nu} = S_{\mathrm{m}}\bigg({\frac{\nu}{\nu_\mathrm{m}}}\bigg)^{\alpha_{\mathrm{t}}}\frac{1-\mathrm{exp}(-\tau_{\mathrm{m}}(\nu/\nu_\mathrm{m})^{\alpha_\mathrm{0}-\alpha_\mathrm{t}})}{1-\mathrm{exp}(-\tau_{\mathrm{m}})},
\end{equation}
where $\tau_{\mathrm{m}} \approx 3/2\bigg(\sqrt{1-\frac{8\alpha_\mathrm{0}}{3\alpha_\mathrm{t}}}-1\bigg)$ is the optical depth at the turnover frequency ($\nu_\mathrm{m}$), 
$S_{\mathrm{m}}$ is the peak flux density corresponding to the flux density at $\nu_\mathrm{m}$, while $\alpha_\mathrm{t}$ and $\alpha_\mathrm{0}$ are the spectral indices for the respective optically thick and thin parts of the SSA spectrum.
The SSA model apparently recognizes curvature in the observed spectrum shown by the peaks in the graphs (for example, the dashed lines in Figure \ref{fig:SSA spectra}).
We assume that the observed spectra result from the superposition of emissions from the steady-state and shocked regions of the jet.
In other words, the observed spectra are composed of the quiescent spectrum and the flaring spectrum.

\begin{figure}
    \centering
    \includegraphics[width=0.45\textwidth]{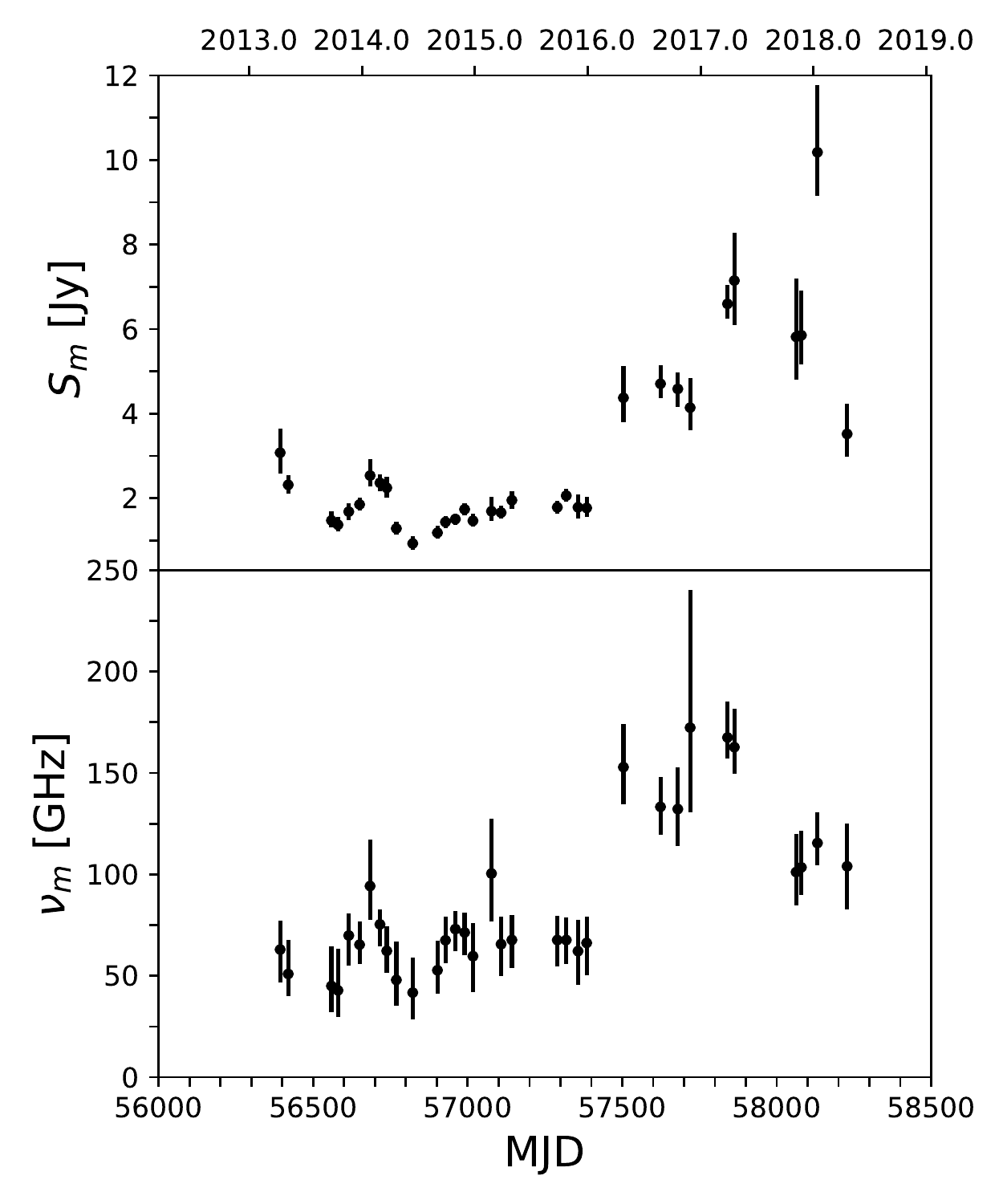}
    \caption{The variation of the peak flux density ($S_\mathrm{m}$) in Jy (top) and of the turnover frequency ($\nu_\mathrm{m}$) in GHz (bottom).}
    \label{fig:Peak Variations}
\end{figure}

Before performing the spectral analysis, we removed the contribution of the quiescent spectrum from the obtained flux densities.
The quiescent spectrum is obtained from \citet{Fromm+2011} as $S(\nu) = c_\mathrm{q}\nu^{\alpha_\mathrm{q}}$, where $c_\mathrm{q}=7.43 \pm 0.65$ and $\alpha_\mathrm{q}=-0.45 \pm 0.04$.
It should be noted that we used the current multi-frequency data from $\sim$2--343~GHz in the period 2012 November--2016 April ($\sim$MJD 56250--57500) to determine the quiescent spectrum by fitting a power law to the median flux densities of the lowest three flux density measurements at each frequency.
However, the determined spectrum was flatter than that found in \citet {Fromm+2011} and yielded higher flux densities than the actual flux measurements at several frequencies (i.e., yielded a negative flux density after quiescent spectrum subtraction), implying that this method is not suitable for estimating the quiescent spectrum.
Therefore, we determined to use the quiescent spectrum obtained from \citet{Fromm+2011}.

We used Eq. \ref{eq:SSA} to make spectral fits on the quiescent-spectrum subtracted data.
Out of the 47 epochs of quasi-simultaneous data, 33 fit results were obtained.
We did not make use of four epochs from 2018 February 19 (MJD 58169), 2018 March 2 (MJD 58180), 2018 March 25 (MJD 58203), and 2018 May 31 (MJD 58270), as the number of flux density measurements was less than the five required in each of these epochs.
We did not make use of a further ten epochs.
On 2017 October 21 (MJD 58048), the data coverage is sparse at $\geq 43$~GHz, which could lead to biased results despite the number of flux densities measurements being sufficient (i.e., more than five).
For the other nine epochs, we found that the spectral shapes showed no curvature, but were instead flat or even inverted.
The flat spectra were found to have a spectral index in the range from 0.06 to 0.33 for five epochs from 2015 January 15 (MJD 57038), 2016 January 13 (MJD 57401), 2016 February 11 (MJD 57430), 2016 March 1 (MJD 57449), and 2018 May 26 (MJD 58265).
The inverted spectra were found to have a spectral index in the range from 0.54 to 0.78 for the four epochs from 2013 March 28 (MJD 56379), 2016 December 28 (MJD 57751), 2017 May 21 (MJD 57895), and 2017 June 17 (MJD 57922).
Figure \ref{fig:SSA spectra} shows the spectral fit results using the single-dish and array data.
The four parameters estimated from the spectral fits are summarized in Table \ref{table: SA - turnover parameters}, while the time evolution of turnover frequency and peak flux density are shown in Figure \ref{fig:Peak Variations}.

\subsection{A Compact Region in Single-dish and VLBI Observations} \label{subsec: compact region}

The beam sizes of the OVRO 40-m radio telescope and the KVN 21-m radio telescope are $\sim$157~arcsec and $\sim$63~arcsec, respectively.
Since those beam sizes are much larger than the beam size (0.12~mas) of the VLBA, the single-dish observations may probe emission regions much larger in size than those probed by the VLBI.
However, the ratio of the mean flux density of the jet components to the mean flux density of the core is 5~{per cent} at 15~GHz.
Similarly, this ratio is 11~{per cent} at 43~GHz.
The core dominance indicates that the variability from single-dish observations could be mainly attributed to the variability in the core region.
In order to confirm this, we compared the size of the emission regions in the single-dish and VLBI observations.

We estimated the size of the variable emission region using the variability of the single-dish data at 15~GHz in Section \ref{subsec:variability parameter} (see, Table \ref{table: flare parameters}).
The VLBI core size was estimated from the VLBA 15~GHz data to be in the range 0.05--0.15~mas with an average of 0.09~mas and a standard deviation of 0.04~mas.
Whereas for the jet components, the average of their sizes is 1.45~mas and its standard deviation is 1.36~mas.
The size of the VLBI core at 43~GHz is in the range 0.01--0.11~mas with a mean of 0.05~mas and a standard deviation of 0.02~mas, which is smaller than the variability size from the 15~GHz single-dish and (slightly) smaller than the VLBI core size at 15~GHz.
The size of the variable emission region is broadly consistent with the size of the VLBI core.
The quantitative agreements between the sizes of the variable emission region and the compact VLBI core indicate their possible association.

\begin{figure}
    \centering
    \includegraphics[width=0.45\textwidth]{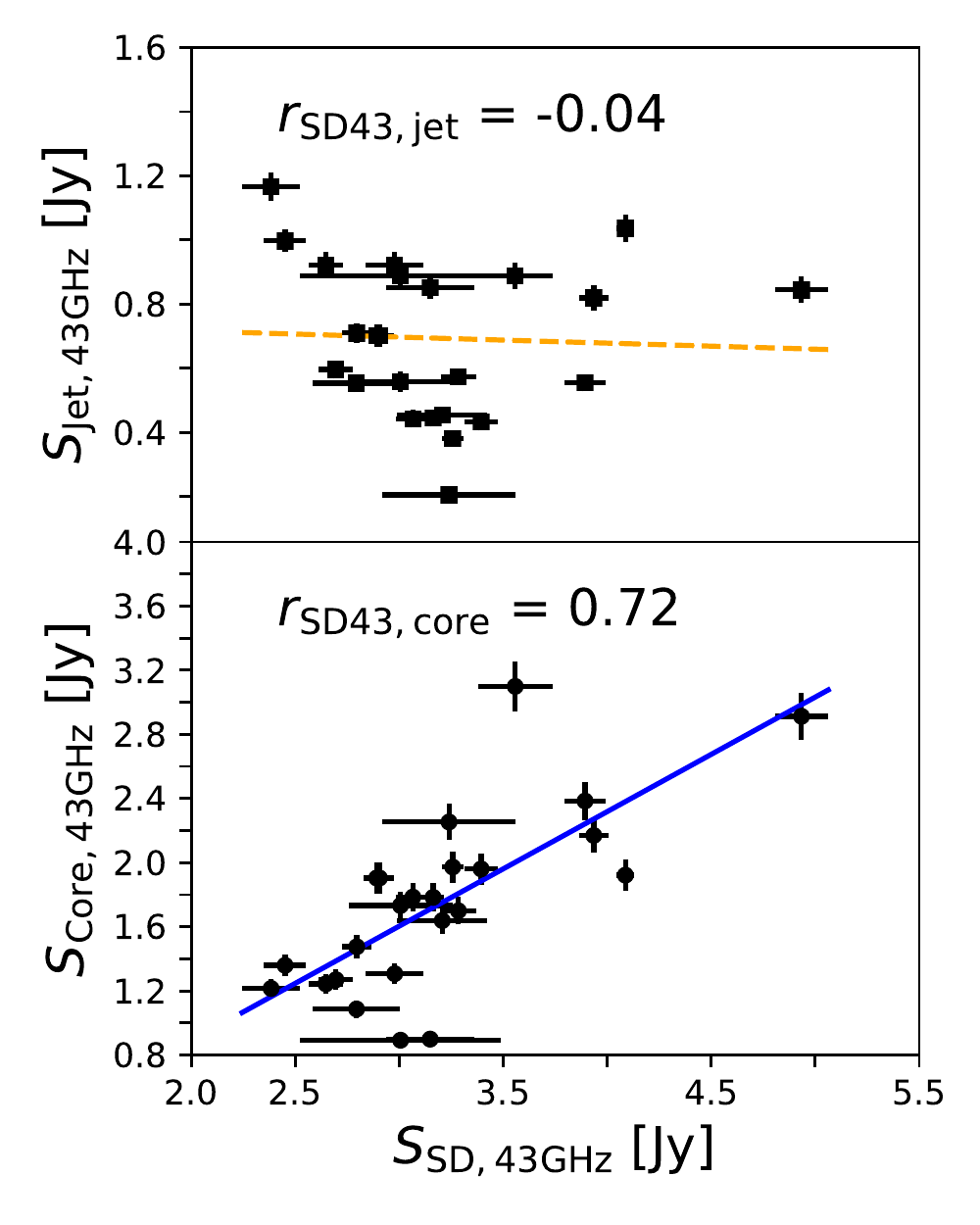}
    \caption{Comparison between the flux densities of the single-dish and the VLBI core/jet components at 43~GHz. The upper row indicates the flux density of jet components $S_\mathrm{jet, 43 GHz}$. The lower row indicates the flux density of core components $S_\mathrm{core, 43 GHz}$. The column shows the flux density of the single-dish. The blue solid line indicates the Pearson correlation between the flux densities of the single-dish and the core component. The orange dashed line indicates the Pearson correlation between the flux densities of the single-dish and the jet components. The Pearson correlation coefficients ($r_\mathrm{SD43, core}$ and $r_\mathrm{SD43, jet}$) are shown in the top left corner of each panel.}
    \label{fig:VLBI vs SD flux}
\end{figure}

We also investigated the Pearson correlation between flux densities of the VLBI core (and jet components) and the single-dish.
Figure \ref{fig:VLBI vs SD flux} shows the results of the Pearson correlation analysis.
We found that the Pearson correlation coefficients are $r_{\rm SD43, core} = 0.72$ for the 43~GHz core flux vs. the 43~GHz single-dish flux and $r_{\rm SD43, jet} = -0.04$ for the 43~GHz jet components flux vs. the 43~GHz single-dish flux, implying that the core flux variability strongly correlates with the single-dish flux variability at 43~GHz.
This confirms that the variability of the radio emissions that were observed may come from the very compact core region.

Therefore, we can posit that the single dish light curve equivalently delivers crucial information relevant to the magnetic field strength on the VLBI core scale.
In the following, we describe the estimation of the magnetic field strength on that scale.

\subsection{Magnetic Fields}    \label{subsec: B-field}

\begin{table*}
\caption{Magnetic Field Strengths of the SSA Region in CTA~102}
\label{table: B-field strengths}
\centering
{
\begin{threeparttable}
\begin{tabular}{cccccc}
\hline
Date & MJD & $b(\alpha)$ & $d_\mathrm{m}$~(mas) & $B_\mathrm{SSA}$~(mG) & $B_\mathrm{eq}$~(mG) \\
(1) & (2) & (3) & (4) & (5) & (6) \\
\hline
2013/4/11  & 56393 & $>3.80$ & $0.034\pm0.001$ & $>0.35$                   & $89 \pm 27$  \\
2013/5/7   & 56419 & $>3.80$ & $0.034\pm0.001$ & $>0.21$                   & $79 \pm 24$  \\
2013/9/24  & 56559 & $>3.80$ & $<0.092$        & --                        & $>29$             \\
2013/10/15 & 56580 & $3.44$  & $<0.092$        & $<12.62$                  & $>28$             \\
2013/11/19 & 56615 & $3.76$  & $0.038\pm0.001$ & $<5.73$                   & $69 \pm 21$  \\
2013/12/24 & 56650 & $3.75$  & $0.057\pm0.001$ & $9.20^{+9.17}_{-7.91}$    & $49 \pm 14$  \\
2014/1/27  & 56684 & $3.26$  & $<0.038$        & $<5.45$                   & $>80$             \\
2014/2/28  & 56716 & $3.37$  & $0.060\pm0.005$ & $12.28^{+9.39}_{-11.25}$  & $52 \pm 16$  \\
2014/3/22  & 56738 & $>3.80$ & $0.060\pm0.005$ & $>5.92$                   & $50 \pm 15$  \\
2014/4/22  & 56769 & $>3.80$ & $0.108\pm0.012$ & $>52.15$                  & $24 \pm 7$    \\
2014/6/13  & 56822 & $>3.80$ & $0.049\pm0.005$ & $>2.18$                   & $43^{+14}_{-13}$  \\
2014/9/1   & 56902 & $>3.80$ & $0.068\pm0.010$ & $>15.31$                  & $36 \pm 12$  \\
2014/9/27  & 56928 & $>3.80$ & $0.068\pm0.010$ & $>35.68$                  & $40 \pm 13$  \\
2014/10/29 & 56960 & $3.75$  & $<0.055$        & $<20.26$                  & $>49$             \\
2014/11/28 & 56990 & $>3.80$ & $<0.055$        & --                        & $>51$             \\
2014/12/25 & 57017 & $3.21$  & $0.046\pm0.005$ & $<8.29$                   & $55 \pm 17$  \\
2015/2/23  & 57077 & $3.76$  & $0.083\pm0.006$ & $<1028.34$                & $37 \pm 11$  \\
2015/3/26  & 57108 & $3.41$  & $0.066\pm0.004$ & $<41.49$                  & $42 \pm 13$  \\
2015/4/30   & 57143 & $3.37$  & $0.071\pm0.010$ & $<44.79$                  & $42 \pm 13$  \\
2015/9/24  & 57290 & $3.61$  & $0.048\pm0.003$ & $<11.24$                  & $57 \pm 17$  \\
2015/10/23 & 57319 & $3.66$  & $0.048\pm0.003$ & $<8.40$                   & $59 \pm 18$  \\
2015/11/30  & 57357 & $3.25$  & $0.057\pm0.002$ & $<15.52$                  & $48 \pm 14$  \\
2015/12/28 & 57385 & $2.81$  & $0.054\pm0.004$ & $<13.41$                  & $51 \pm 16$  \\
2016/4/25  & 57504 & $>3.80$ & $<0.027$        & --                        & $>136$            \\
2016/8/23  & 57624 & $3.62$ & $0.071\pm0.010$ & $<44.80$                  & $>98$            \\
2016/10/18 & 57680 & $>3.80$ & $<0.068$        & --                        & $>61$             \\
2016/11/27 & 57720 & $3.48$  & $0.063\pm0.006$ & $<962.19$                 & $66^{+21}_{-20}$  \\
2017/3/28  & 57841 & $3.60$  & $<0.043$        & $<25.58$                  & $>103$            \\
2017/4/20  & 57863 & $>3.80$ & $0.041\pm0.002$ & $>16.50$                  & $109 \pm 33$ \\
2017/11/4  & 58063 & $3.79$  & $<0.048$        & $<3.99$                   & $>85$             \\
2017/11/22 & 58080 & $3.46$  & $<0.048$        & $<4.03$                   & $>86$             \\
2018/1/13  & 58132 & $3.54$  & $0.102\pm0.002$ & $50.97^{+42.80}_{-33.92}$ & $53 \pm 16$  \\
2018/4/19  & 58228 & $>3.80$ & $0.049\pm0.002$ & $>14.10$                  & $72 \pm 22$ \\
\hline
\end{tabular}
\begin{tablenotes}
\item Note. Column designation: (1) observing date in year/month/day, (2) observing date in modified Julian date, (3) parameter depending on the optically thin spectral index (see the text), (4) size of emission region at turnover frequency in mas, (5) magnetic field strength of SSA region in mG, and (6) magnetic field strength in equipartition conditions in mG
\end{tablenotes}
\end{threeparttable}
}
\end{table*}

The magnetic field strength in an SSA region is estimated through the synchrotron self-absorbed spectra.
The magnetic field in the SSA region $B_\mathrm{SSA}$ \citep{Marscher+1983} can be expressed as:
\begin{equation}    \label{eq:B_SSA}
    B_\mathrm{SSA} = 10^{-5}b(\alpha)S_\mathrm{m}^{-2}d_\mathrm{m}^4\nu_\mathrm{m}^5{\bigg(\frac{\delta}{1+z}\bigg)}^{-1}~[G],
\end{equation}
where $b(\alpha)$ is a dimensionless parameter depending on the optically thin spectral index \citep[see Table 1 in][]{Marscher+1983}, $S_{\mathrm{m}}$ is the peak flux density in Jy, $d_\mathrm{m}$ is the angular size of the VLBI core at the turnover frequency in mas (assuming the SSA region is dominated by the VLBI core emission region), $z$ is the redshift of the source, and $\delta$ is the Doppler factor.
In contrast to \citet{Marscher+1983}, which is related to moving features, we adopt the $(\frac{\delta}{1+z})$ factor being raised to the $-1$ power rather than $+1$ power since we consider the VLBI core to be in a steady state rather than evolving with time, in other words, it is a stationary component \citep[e.g.,][]{LeeJW+2017,Algaba+2018b}.

The parameter $b(\alpha)$ which depends on the optically thin spectral index $\alpha_\mathrm{0}$ is in the range 1.8--3.8 \citep[see, Table 1 in][]{Marscher+1983}.
Otherwise, we take 1.8 as the upper limit and 3.8 as the lower limit.
We assume that the size of the SSA region is close to the 43~GHz VLBI core size, $d_\mathrm{m} \simeq d_{\rm 43~GHz}$, as summarized in Table \ref{table: B-field strengths}.
The core component at 43~GHz is considered to be unresolved when the core size $d_\mathrm{43~GHz}$ is smaller than its minimum resolvable size $d_{\rm min}$ \citep[see,][for more details]{LeeSS+2016b}.
We take the minimum resolvable size as an upper limit of the core size if $d_\mathrm{43~GHz} < d_{\rm min}$.
The time difference between the epochs of the VLBA data and the epochs when turnover frequencies are estimated is about 10~days.
Furthermore, we need a Doppler factor (i.e., a Doppler factor in the SSA region) to calculate magnetic field strengths in the SSA region.
In Section \ref{subsec:variability parameter}, we used $\delta = 17 \pm 3$ (obtained from the 15~GHz VLBI observations) to estimate the characteristic size ($d_{\rm var}$) of the flux density variability at 15~GHz (i.e., the size of the variable emission region at 15~GHz).
We note that the SSA region is most likely to be closer to the 43~GHz core than to the 15~GHz one since most turnover frequencies are $\nu_{\rm m} \geq 43~{\rm GHz}$.
\cite{Jorstad+2017} measured the motions of the moving jet component nearest to the VLBI core at 43~GHz and computed the Doppler factor of $\delta = 30.8 \pm 12.9$.
We take this Doppler factor to derive the magnetic field strengths.

\begin{figure}
    \centering
    \includegraphics[width=0.45\textwidth]{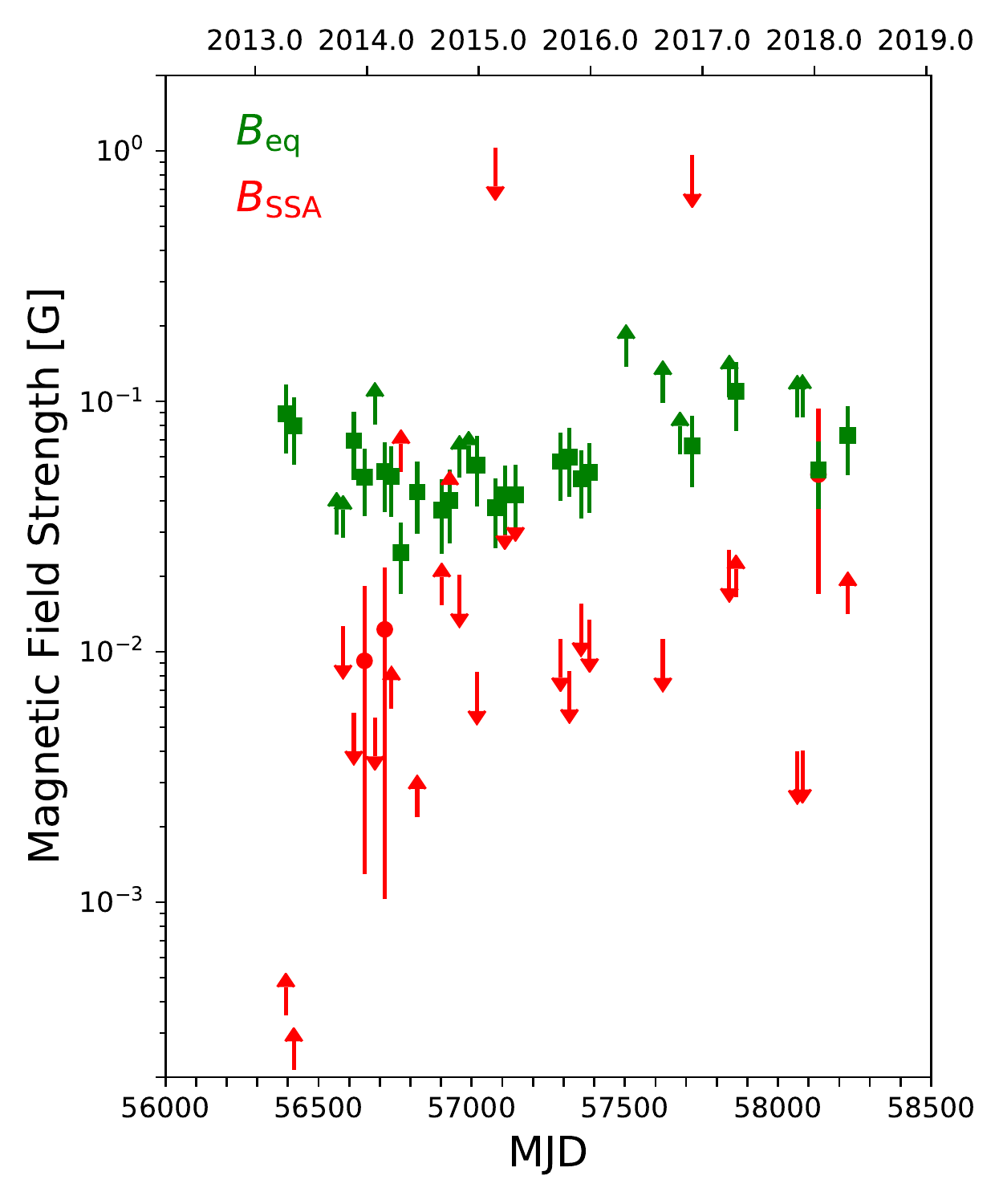}
    \caption{Magnetic field strengths of CTA~102. The magnetic fields in equipartition conditions ($B_\mathrm{eq}$) are shown by the green squares while the magnetic fields in SSA regions ($B_\mathrm{SSA}$) are shown by red points. The arrows with different colors indicate the upper and lower limits of $B_\mathrm{eq}$ (green) and $B_\mathrm{SSA}$ (red) following the directions of those arrows.}
    \label{fig:B-fields}
\end{figure}

Taking into account all the parameters in Eq. \ref{eq:B_SSA}, the SSA magnetic field strength can be estimated.
We take the SSA magnetic field strength $B_{\rm SSA}$ as the upper limit or the lower limit depending on two parameters $b(\alpha)$ and $d_\mathrm{m}$ in Eq. \ref{eq:B_SSA}.
If the value of $b(\alpha)$ is taken as the lower limit while the value of $d_\mathrm{m}$ is taken as the upper limit, we rule out the SSA magnetic field strength.
Furthermore, we take $B_{\rm SSA}$ as the upper limit if the derivation is lower than its uncertainty in the lower bound.
Then, the upper limit for $B_{\rm SSA}$ is regarded as the sum of its derivation and the uncertainty in the upper bound.
In this manner, three derivations of the SSA magnetic field strength are $B_\mathrm{SSA}=9.20^{+9.17}_{-7.91}$~mG,  $B_\mathrm{SSA}=12.28^{+9.39}_{-11.25}$~mG, and $B_\mathrm{SSA}=50.97^{+42.80}_{-33.92}$~mG for the three epochs on 2013 December 24 (MJD 56650), 2014 February 27 (MJD 56716), and 2018 January 13 (MJD 58132), respectively.
The upper or lower limits for $B_\mathrm{SSA}$ were also obtained for all the other epochs.
The results are summarized in Table \ref{table: B-field strengths} and Figure \ref{fig:B-fields}.

We can also calculate the magnetic field strength assuming equipartition between the energy densities of the radiating particles and the magnetic fields in the SSA region.
The magnetic field strength under the equipartition condition, $B_\mathrm{eq}$, can be described as shown in \citep{Kataoka&Stawarz+2005}:
\begin{equation}    \label{eq:B_eq}
\begin{aligned}
    B_\mathrm{eq} = 1.23 \times 10^{-4} \eta^{2/7}{(1+z)}^{11/7}{\bigg(\frac{D_\mathrm{L}}{100~\mathrm{Mpc}}\bigg)}^{-2/7}\\
    \times {\bigg(\frac{\nu_\mathrm{m}}{5~\mathrm{GHz}}\bigg)}^{1/7}{\bigg(\frac{S_\mathrm{m}}{100~\mathrm{mJy}}\bigg)}^{2/7}{\bigg(\frac{10^3d_\mathrm{m}}{0.3~{\rm arcsec}}\bigg)}^{-6/7}\delta^{-5/7}~[G],
\end{aligned}
\end{equation}
where $\eta$ is the ratio of the energy density carried by protons and electrons to the energy density of the electrons (i.e., $\eta=$ 1 and 1836 for the leptonic and hadronic jets, respectively).
Here we assume $\eta\sim1$.
Other parameters in Eq. \ref{eq:B_eq} are identical to those in Eq. \ref{eq:B_SSA}.
We take the equipartition magnetic field strength as the lower limit depending on the core size at the turnover frequency $d_\mathrm{m}$.
The equipartition magnetic field strengths were measured for 22 epochs and their lower limits were obtained for 10 epochs.
The magnetic field strength derivations under the equipartition condition, $B_\mathrm{eq}$, are in the range 24--109~mG.
The estimated $B_\mathrm{eq}$ are summarized in Table~\ref{table: B-field strengths} and described in Figure~\ref{fig:B-fields}.

The comparison of the magnetic field strengths between $B_{\rm SSA}$ and $B_{\rm eq}$ may help us understand the physical characteristics in the inner region of the jet.
In a total of 16 epochs, we found the significant difference between $B_{\rm SSA}$ and $B_{\rm eq}$ considering their uncertainties, i.e., $B_{\rm SSA}/B_{\rm eq} \ll 1$ or $B_{\rm SSA}/B_{\rm eq} \gg 1$.
Comparing the derivations of $B_{\rm SSA}$ to those of $B_{\rm eq}$, we found that $B_{\rm SSA}$ is smaller than $B_{\rm eq}$ by a factor of 4--5 for two epochs on 2013 December 24 (MJD 56650) and 2014 February 27 (MJD 56716).
We also found that the upper limits of $B_{\rm SSA}$ are smaller than the derivations or lower limits of $B_{\rm eq}$ in 13 epochs (see Figure \ref{fig:B-fields} and Table \ref{table: B-field strengths}).
The ratio $B_{\rm SSA} / B_{\rm eq} \ll 1$ indicates that the SSA region is dominated by particle energy.
On 2014 April 22 (MJD 56769), the lower limit of $B_{\rm SSA}$ is larger than the derivation of $B_{\rm eq}$, implying that the SSA region is dominated by magnetic field energy.
However, the derivation of $B_{\rm SSA}$ is comparable to that of $B_{\rm eq}$ on 2018 January 13 (MJD 58132), which indicates that the SSA region is near equipartition between magnetic field energy and particle energy.
The comparison between the magnetic field strengths of $B_{\rm SSA}$ and $B_{\rm eq}$ (i.e., dominating energy in the jet regions) enables us to discuss the physical properties of the relativistic jets (including e.g., the particle acceleration mechanism).
For example, if the jet is dominated by magnetic field energy, magnetic reconnection can be the particle acceleration mechanism, while if the jet is dominated by particle energy, shock acceleration can be that mechanism \citep[see,][and references therein]{Hovatta+2019}.
This suggests that we can present possibilities of flare origins.
In the following sections, we further discuss these results.

\section{Discussions}    \label{sec:discussion}

\subsection{Variability Doppler Factor in Equipartition Region}
In Section \ref{subsec: B-field}, the derivation of the magnetic field strengths ($B_{\rm SSA} \approx B_{\rm eq}$) for 2018 January (MJD 58132) implies that the SSA region in the jet of CTA~102 is under equipartition conditions at that time.
In Section \ref{subsec:decomposition}, we identified seven flares in the 15~GHz light curve.
Based on three parameters for each flare, namely the maximum amplitude, the time of the maximum amplitude, and the variability time-scale, we estimated the variability brightness temperature ($T_\mathrm{b, var}$) in Section \ref{subsec:variability parameter}.
In order to derive the variability Doppler factor, not only the variability brightness temperature but also the intrinsic brightness temperature are required.
It should be noted that the epoch of 2018 January (MJD 58132), which is under equipartition conditions, corresponds to the period of Flare~7.
We assume that the source becomes close to equipartition between the energy densities of the magnetic field and the particles while a flare is underway \citep{Readhead+1994, Lahteenmaki+1999}.
Under the equipartition assumption, we can substitute the intrinsic brightness temperature for the equipartition brightness temperature ($T_\mathrm{eq}$).
For the equipartition brightness temperature, we use $T_\mathrm{eq} = 5 \times 10^{10}$~K \citep{Readhead+1994}.
Then, we compute the variability Doppler factor $\delta_\mathrm{var}$ \citep{Lahteenmaki&Valtaoja+1999}, defined as:
\begin{equation}
    \delta_\mathrm{var} = (1+z)\sqrt[3]{\frac{T_\mathrm{b, var}}{T_\mathrm{eq}}}.
\end{equation}
The variability Doppler factor for Flare~7 is $\delta_\mathrm{var} = 17.79^{+1.35}_{-0.88}$.

We compare our estimate of $\delta_\mathrm{var}$ with the variability Doppler factor estimated in \citet{Liodakis+2018}.
\citet{Liodakis+2018} computed the variability Doppler factor of CTA~102 and found that the variability Doppler factor is $\delta = 13.39^{+4.16}_{-5.20}$ using the same 15~GHz data from similar periods.
\citet{Liodakis+2018} used $T_\mathrm{eq} = (2.78 \pm 0.72) \times 10^{11}$~K, whereas we used $T_\mathrm{eq} = 5 \times 10^{10}$~K.
If we used $T_\mathrm{eq} = (2.78 \pm 0.72) \times 10^{11}$~K instead, as \citet{Liodakis+2018} adopted, we obtain a variability Doppler factor of $\delta_\mathrm{var}=10.04^{+1.16}_{-1.00}$.
Thus, our Doppler factors are consistent with those in \citet{Liodakis+2018}.

\subsection{Magnetic Field Strengths of Blazars}
For several blazars, the magnetic field strengths have been estimated using a similar method to one in this paper: OJ 287 \citep{LeeJW+2020}, 0716+714 \citep{LeeJW+2017}, 1156+295 \citep{Kang+2021}, and 1633+382 \citep{Algaba+2018b}, which are in the range of redshift $z=0.127$--$1.814$.
The ranges of the SSA magnetic field strengths of those blazars (called $B_\mathrm{SSA, source}$) are as follows: $B_\mathrm{SSA, 0716+714} \sim 0.04$--$4.58$~mG (all upper limits), $B_\mathrm{SSA, OJ287}=0.16$--$0.26$~mG, $B_\mathrm{SSA, 1156+295}=1$--$99$~mG, and $B_\mathrm{SSA, 1633+382}=0.01$--$0.19$~mG.
The equipartition magnetic field strengths (called $B_\mathrm{eq, source}$) were also computed for those blazars, except for 0716+714.
The ranges are $B_\mathrm{eq, OJ287}=0.95$--$1.93$~mG, $B_\mathrm{eq, 1156+295}=100$--$254$~mG, and $B_\mathrm{eq, 1633+382}=117$--$1146$~mG.
We found that for those blazars, $B_\mathrm{eq}$ is usually larger than $B_\mathrm{SSA}$, which is similar to our results.

\begin{table*}
\caption{Lists of parameters obtained from the magnetic field strength ratio}
\label{table: Parameters obtained from B-field ratio}
\centering
{
\begin{threeparttable}
\begin{tabular}{cccccccc}
\hline
Date & MJD & $u_{\rm p}/u_{\rm B}$ & $u/u_{\rm eq}$ & $T_{\rm b,int}~({\rm 10^{11} K})$ & $P_{\rm B}~({\rm erg \cdot s^{-1}})$ & $P_{\rm p}~({\rm erg \cdot s^{-1}})$ & $P_{\rm syn}~({\rm erg \cdot s^{-1}})$ \\
(1) & (2) & (3) & (4) & (5) & (6) & (7) & (8) \\
\hline
2013/10/15 & 56580                                         & $> 32$ & $> 3.22$ & $< 0.75$ & $< 3.17e+44$ & $> 1.01e+46$ & $< 2.26e+42$  \\
2013/11/19 & 56615                                         & $> 4.09 e+4$ & $> 138$ & $< 1.74$ & $< 1.11e+43$ & $> 4.56e+47$ & $< 2.72e+42$  \\
2013/12/24 & 56650                                         & $< 7.03 e+3$ & $< 74$ & $1.16^{+0.60}_{-0.53}$ & $< 1.93e+44$ & $< 3.00e+47$ & $< 8.00e+42$  \\
2014/1/27  & 56684                                         & $> 9.27e+4$ & $> 213$ & $< 1.92$ & $< 1.01e+43$ & $> 9.36e+47$ & $< 4.76e+42$  \\
2014/2/28  & 56716                                         & $< 2.15e+3$ & $< 37$ & $1.03^{+0.43}_{-0.50}$ & $< 3.24e+44$ & $< 1.91e+47$ & $< 1.13e+43$  \\
2014/4/22  & 56769                                         & $< 4.29e-2$ & $< 2.29$ & $> 0.34$ & $> 7.46e+45$ & $< 3.20e+44$ & $< 1.78e+42$  \\
2014/10/29 & 56960                                         & $> 44$ & $> 3.81$ & $< 0.78$ & $< 2.92e+44$ & $> 1.30e+46$ & $< 2.58e+42$  \\
2014/12/25 & 57017                                         & $> 3.23e+3$ & $> 36$ & $< 1.29$ & $< 3.41e+43$ & $> 1.10e+47$ & $< 2.76e+42$  \\
2015/9/24  & 57290                                         & $> 1.03e+3$ & $> 19$ & $< 1.13$ & $< 6.84e+43$ & $> 7.04e+46$ & $< 3.02e+42$  \\
2015/10/23 & 57319                                         & $> 4.22e+3$ & $> 42$ & $< 1.34$ & $< 3.82e+43$ & $> 1.61e+47$ & $< 3.41e+42$  \\
2015/11/30 & 57357                                         & $> 132$ & $> 6.68$ & $< 0.89$ & $< 1.84e+44$ & $> 2.43e+46$ & $< 3.19e+42$  \\
2015/12/28 & 57385                                         & $> 316$ & $> 11$ & $< 0.98$ & $< 1.23e+44$ & $> 3.89e+46$ & $< 3.17e+42$  \\
2016/8/23  & 57624                                         & $> 1.01e+4$ & $> 66$ & $< 1.48$ & $< 4.74e+43$ & $> 4.79e+47$ & $< 8.49e+42$  \\
2017/3/28  & 57841                                         & $> 379$ & $> 12$ & $< 1.01$ & $< 2.84e+44$ & $> 1.08e+47$ & $< 1.16e+43$ \\
2017/11/4  & 58063                                         & $> 4.63e+5$ & $> 499$ & $< 2.32$ & $< 8.63e+42$ & $> 3.99e+48$ & $< 9.13e+42$  \\
2017/11/22 & 58080                                         & $> 4.55e+5$ & $> 495$ & $< 2.32$ & $< 8.78e+42$ & $> 4.00e+48$ & $< 1.01e+43$ \\
2018/1/13  & 58132                                         & $< 5.75$ & $1.01^{+0.29}_{-0.23}$ & $0.51^{+0.23}_{-0.19}$ & $< 1.70e+46$ & $< 2.51e+46$ & $< 4.68e+43$ \\
\hline
\end{tabular}
\begin{tablenotes}
\item Note. Column designation: (1) observing date in year/month/day, (2) observing date in modified Julian date, (3) ratio of particle energy density to magnetic field energy density, (4) ratio of total energy density to total energy density in equipartition state, (5) intrinsic brightness temperature in $10^{11}$~K, (6) jet power for magnetic field in ${\rm erg \cdot s^{-1}}$, (7) jet power for particles in ${\rm erg \cdot s^{-1}}$, and (8) jet power for synchrotron radiation in ${\rm erg \cdot s^{-1}}$
\end{tablenotes}
\end{threeparttable}
}
\end{table*}

\subsection{Physical Properties of SSA Regions}  \label{sec: SSA regions}
In this section, we estimate physical parameters (e.g., intrinsic brightness temperature and jet power) by using the magnetic field strength ratio (i.e., $B_{\rm SSA}/B_{\rm eq}$).
In order to do this, we used the magnetic field strengths for 17 epochs when we found the dominant source of the energy in the SSA regions (see Section \ref{subsec: B-field}).
Note that in the case of using the derivations of $B_{\rm SSA}$ and $B_{\rm eq}$, we present the parameters as upper limits if the uncertainties of the parameters are larger than their derivations.
For the case where the limits of $B_{\rm SSA}$ and $B_{\rm eq}$ were used, we exhibit the parameters as the lower or upper limits, after considering the magnetic field strength limits in the equations of the parameters.
We summarize the parameters in Table \ref{table: Parameters obtained from B-field ratio}.

\subsubsection{Intrinsic Brightness Temperatures in SSA Regions}
The energy of the source (magnetic field energy plus particle energy) can be at its minimum when the source is in equipartition, i.e., when it is in an approximate balance between the magnetic field energy density $u_{\rm B}$ and the particle energy density $u_{\rm p}$, hence when $u_{\rm p}/u_{\rm B} \approx 1$ \citep{Longair+1994, Condon&Ransom+2016}.
One should find the ratio of the energy densities $u_{\rm p}/u_{\rm B}$ is much larger than unity if SSA regions in a relativistic jet are dominated by particle energy.
We assumed that the jet of the source consists of an electron-positron plasma \citep{Readhead+1994}.
Following \citet{Readhead+1994}, $B/B_\mathrm{eq}=(T^{\prime}_\mathrm{eq}/T^{\prime}_\mathrm{b, int})^2$, where B is the measured magnetic field strength (here, we regard $B$ as $B_\mathrm{SSA}$), $T^{\prime}_\mathrm{eq}$ and $T^{\prime}_\mathrm{b, int}$ are the equipartition and intrinsic brightness temperatures in the rest frame of the observer, respectively.
The ratio of the total energy density to its equipartition value is given by 
\begin{equation}
    \frac{u}{u_\mathrm{eq}} = \frac{1}{2}\bigg(\frac{T^{\prime}_\mathrm{eq}}{T^{\prime}_\mathrm{b,int}}\bigg)^{4}\bigg[1+\bigg(\frac{T^{\prime}_\mathrm{eq}}{T^{\prime}_\mathrm{b,int}}\bigg)^{-17/2}\bigg],
\end{equation}
where $u$ is the total energy density and $u_\mathrm{eq}$ is the total energy density in the equipartition state.
Moreover, the ratio of the particle energy density $u_\mathrm{p}$ to the magnetic field energy density $u_\mathrm{B}$ is given by $u_\mathrm{p}/u_\mathrm{B}=({T^{\prime}_\mathrm{eq}/T^{\prime}_\mathrm{b,int}})^{-17/2}$.
In order to estimate the energy density ratios ($u/u_\mathrm{eq}$ and $u_\mathrm{p}/u_\mathrm{B}$), we used the ratio of magnetic field strengths $B/B_\mathrm{eq}$ as follows:
\begin{equation}
    \frac{u}{u_\mathrm{eq}} = \frac{1}{2}\bigg(\frac{B}{B_\mathrm{eq}}\bigg)^{2}\bigg[1+\bigg(\frac{B}{B_\mathrm{eq}}\bigg)^{-17/4}\bigg], \\
\end{equation}

\begin{equation}
    \frac{u_\mathrm{p}}{u_\mathrm{B}}=\bigg(\frac{B}{B_\mathrm{eq}}\bigg)^{-17/4}. \\
\end{equation}

The energy density ratios are $u/u_{\rm eq} < 2.29$ and $u_{\rm p}/u_{\rm B} < 4.29 \times 10^{-2}$ for the magnetically dominated region.
For the particle-energy-dominated region, $u/u_{\rm eq}$ and $u_{\rm p}/u_{\rm B}$ are at least the order of $10^{0}$ and $10^{1}$, but they are $>$$\sim$$10^2$ and $>$$\sim$$10^5$ at extreme cases.
In the case of equipartition, $u/u_{\rm eq} \approx 1$ and $u_{\rm p}/u_{\rm B} < 5.75$.
According to several observational and theoretical expectations, the energy density ratio should be $u_{\rm p}/u_{\rm B} \gtrsim 10^5$ for the departure from the equipartition state due to the injection or acceleration of particles at the jet base \citep{Readhead+1994, Homan+2006, Singal+2009}.
We found that several epochs when the SSA regions are dominated by particle energy agree well with the condition to be out of the equipartition state.

The equipartition brightness temperature in the rest frame of the observer can be obtained as $T^{\prime}_\mathrm{eq} = \delta T_\mathrm{eq} = 1.54 \times 10^{12}$~K, using $T_\mathrm{eq}=5 \times 10^{10}$~K \citep{Readhead+1994} and $\delta = 30.8 \pm 12.9$ \citep{Jorstad+2017}.
The intrinsic brightness temperatures in the source frame are $T_{\rm b, int} <$ $\sim$$10^{11}$ in K.
We found that $T_{\rm b, int}$ is lower than the inverse Compton limit $\sim 10^{12}$~K, which implies that SSA regions in the jet of CTA~102 are not far out of equipartition \citep{Kellermann&Pauliny-Toth+1969, Readhead+1994, Homan+2006, Singal+2009}.
This indicates that the inverse Compton catastrophe, where the energy loss due to inverse Compton scattering decreases the brightness temperature of radio sources $> 10^{12}$~K to lower than this \citep{Kellermann&Pauliny-Toth+1969}, does not apply to the SSA regions.

\subsubsection{Jet Powers in SSA Regions}   \label{subsubsec:jet power}

Based on the physical parameters (e.g., magnetic field strengths, SSA spectral properties) in SSA regions, we can estimate the jet powers of CTA~102.
To estimate the jet powers, we assume that the emissions from radio to $\gamma$-rays can be interpreted in terms of the one-zone homogeneous model \citep{Celotti&Ghisellini+2008}.
In this model, two components, low-energy (radiated by the synchrotron process) and high-energy (radiated by the inverse-Compton scattering) components, are produced by a single relativistic lepton population in the source spectrum.
The jet power carried by relativistic particles (e.g., electrons), the magnetic field, and radiation (synchrotron radiation and inverse Compton scattered radiation) is given by
\begin{equation}    \label{eq:jet power - B}
    P_i = \pi R^2 \Gamma^2 \beta c u_i~[{\rm erg \cdot s^{-1}}],
\end{equation}
where $R$ is the linear radius of the emission region in cm, $\Gamma$ is the bulk Lorentz factor, $\beta c$ is the velocity of the emitting plasma in ${\rm cm \cdot s^{-1}}$, and $u_i$ is the energy density of the $i$ component in ${\rm erg \cdot cm^{-3}}$ ($i$ denotes relativistic particles, the magnetic field, and radiation).
Since the angular size of the SSA region is $d_{\rm m}$, the linear radius is $R = (d_{\rm m}/2)[D_{\rm L}/(1+z)^2]$ \citep[see, e.g., ][]{Celotti&Fabian+1993, Algaba+2018b}.
We assume that $\Gamma \sim 20$ and $\beta c \sim c$.
The energy densities are $u_{\rm p}$ for relativistic particles, $u_{\rm B}$ for the magnetic field, and $u_{\rm rad}$ for radiation.

Taking into account these parameters, we estimate the power of each component contained in the jet.
The magnetic field energy density, $u_{\rm B} = B^2_{\rm SSA}/8 \pi$, enables us to estimate the jet power for the SSA magnetic field $P_{\rm B}$.
Since we already obtained the ratio of $u_{\rm p}/u_{\rm B}$, the jet power for relativistic particles $P_{\rm p}$ can be estimated.
The total jet power can be estimated as $P_{\rm jet} = P_{\rm B} + P_{\rm p}$.
The radiation energy density is measured as $u_{\rm rad} = L/4\pi R^2 c$, where $L$ is the luminosity in ${\rm erg \cdot s^{-1}}$, which is assumed to be isotropic \citep{Ghisellini+2013}.
The synchrotron luminosity at a monochromatic frequency \citep{Pacholczyk+1970} is $L_{{\rm syn}, \nu} = 4 \pi D_{\rm L}^2 S_{\nu}$, where $S_{\nu}$ is the flux density in the SSA region in Jy (Eq. \ref{eq:SSA}).
The synchrotron luminosity in the rest frame of the observer $L^{\prime}_{\rm syn}$ is estimated from the flux density integrated over a range of frequencies (2--340~GHz).
The synchrotron luminosity in the source frame is $L_{\rm syn} = L^{\prime}_{\rm syn}/\delta^4$.
Using $L \approx L_{\rm syn}$ for the radiation energy density, the synchrotron radiative jet power $P_\mathrm{syn}$ can be obtained.
The central black hole mass of CTA~102 gives the Eddington power of the black hole of $\sim 1.1 \times 10^{47}~{\rm erg \cdot s^{-1}}$ \citep{Zamaninasab+2014, Zacharias+2017, Prince+2018}.
We found that all jet powers, except for $P_{\rm p}$, are plausible considering the Eddington power.
The jet power for the magnetic field is $P_{\rm B} <$ $\sim$$10^{46}~{\rm erg \cdot s^{-1}}$, except for $P_{\rm B} > 7.46 \times 10^{45}~{\rm erg \cdot s^{-1}}$ on 2014 April 22 (note that the SSA region is magnetically dominated in this epoch).
All synchrotron radiative jet powers are $P_{\rm syn} <$ $\sim$$10^{43}~{\rm erg \cdot s^{-1}}$.
For epochs when $P_{\rm p}$ is higher than the Eddington power, the particle energy density is $\sim$$10^4$ times larger than the magnetic field energy density.
This suggests that strong activities in the relativistic jet increased the particle energy density and resulted in the jet power exceeding the Eddington power temporarily.

Previous studies \citep[e.g.,][]{Celotti&Ghisellini+2008, Kang+2014, Zheng+2017, Gasparyan+2018, Prince+2018} estimate the jet power of components (i.e., magnetic field, electrons, and radiation) by fitting the one-zone SED model using optical to $\gamma$-ray emissions.
We now compare the jet powers from CTA~102 that we estimated and those from the same source presented in the other studies.
The jet powers from previous studies are in the following: $P_{\rm B} \sim 10^{45-46}~{\rm erg \cdot s^{-1}}$ \citep{Celotti&Ghisellini+2008, Kang+2014, Gasparyan+2018, Prince+2018}, $P_{\rm p} \sim 10^{44-45}~{\rm erg \cdot s^{-1}}$ \citep{Celotti&Ghisellini+2008, Kang+2014, Gasparyan+2018, Prince+2018}, $P_{\rm jet} \sim 10^{46-47}~{\rm erg \cdot s^{-1}}$ \citep{Celotti&Ghisellini+2008, Kang+2014, Zheng+2017, Gasparyan+2018, Prince+2018}, and $P_{\rm syn} \sim 10^{44}~{\rm erg \cdot s^{-1}}$ \citep{Celotti&Ghisellini+2008, Kang+2014}.
Our estimates of $P_{\rm B}$ and $P_{\rm p}$ in the magnetically dominated and equipartition regions are consistent with those from previous studies.
However, for the particle-energy-dominated region, $P_{\rm B}$ is lower than that from previous studies, while $P_{\rm p}$ is higher than that from previous studies.
Our estimates of $P_{\rm syn}$ are lower than those from previous studies regardless of the energy dominance in SSA regions.
We found that the magnetic field strengths of the emission regions reported in previous studies \citep{Celotti&Ghisellini+2008, Kang+2014, Gasparyan+2018, Prince+2018} are in the range 1.0--6.1 G, which is much higher than our estimates of $B_{\rm SSA} \sim 9-50$~mG and $B_{\rm eq} \sim 24-109$~mG.
This implies that the difference in jet powers between previous studies and ours can be attributed to the large difference in the magnetic field strength being derived for different regions of the jet (i.e., the upstream jet for the SED method and the downstream jet for our study).

Flares at high energies (from optical to $\gamma$-ray) as well as at radio energies were observed from CTA~102 during the period our data covers \citep[e.g.,][]{Raiteri+2017, D'Ammando+2019}.
In the period 2016 November--2017 February, the $\gamma$-ray flaring activity was observed to have a peak luminosity of $L_{\gamma} \sim 10^{50}~{\rm erg \cdot s^{-1}}$ \citep{D'Ammando+2019}.
Using the $\gamma$-ray luminosity, the inverse-Compton radiative jet power is $P_{\rm IC} \sim 10^{46}~{\rm erg \cdot s^{-1}}$ if the $\gamma$-ray emission dominates the inverse-Compton scattered emission.
With the synchrotron radiative jet powers of $P_{\rm syn} <$ $\sim$$10^{43}~{\rm erg \cdot s^{-1}}$, the total (synchrotron plus inverse-Compton) radiative jet power is $P_{\rm tot} \sim 10^{46}~{\rm erg \cdot s^{-1}}$, which is broadly consistent with the total jet power.

\subsubsection{Radio Flares in SSA Regions}

\begin{figure*}
    \centering
    \includegraphics[width=1.0\textwidth]{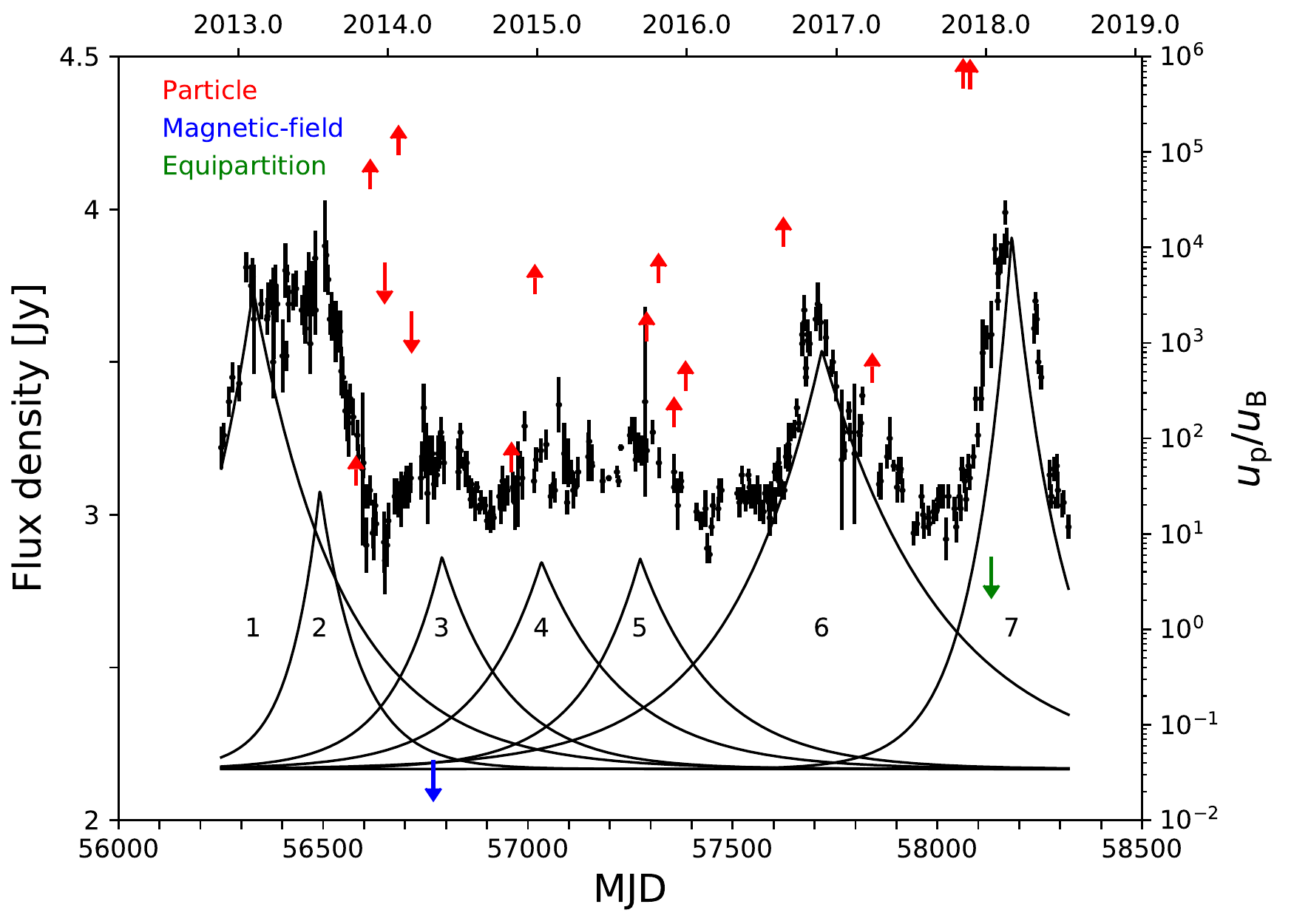}
    \caption{The light curve from CTA~102 at 15~GHz decomposed into seven exponential flares (for more details, see Figure \ref{fig:decomp result}). The arrows are the upper and lower limits of the ratio of the energy densities. The colors indicate the dominant source of jet energy.}
    \label{fig: flares vs u_ratio}
\end{figure*}

The parameter $B_\mathrm{eq}$ indicates the magnetic field strength under equipartition between the energy density of relativistic particles ($u_\mathrm{p}$) and the energy density of the magnetic fields ($u_\mathrm{B}$).
This implies that $B_\mathrm{eq}$ can indicate the particle energy budget ($u_\mathrm{p} \approx u_\mathrm{B} = B_\mathrm{eq}^{2} / 8\pi$).
We found that $B_\mathrm{eq}$ hit a maximum of 109~mG in 2017 April (MJD 57863), increasing from 48~mG in 2015 November (MJD 57357).
Close to the epoch when $B_\mathrm{eq}$ reaches its the maximum, Flare~6 occurs from 2016 April--2017 February (MJD 57500-57800).
Moreover, some active states are seen at $>$ 86~GHz after $\sim$2016 April ($\sim$MJD 57500), as presented in Figure~\ref{fig:light curves}.
Even though we cannot obtain the derivation of the $B_\mathrm{SSA}$ during Flare~6, the large $B_\mathrm{eq}$ while Flare~6 was occurring indicates that Flare~6 (e.g., large synchrotron emissivity) may be related to the increase in the particle energy density $u_{\rm p}$.
The increase in the particle energy density could be attributed to an increased number density of accelerated synchrotron-emitting particles (e.g., electrons).
During the period 2017 September--2018 July (MJD 58000-58300) while Flare~7 occurred, we also found that $B_\mathrm{eq}$ increased from 53~mG to 72~mG.
The fact that the outbreak of Flare~7 and this increase in $B_\mathrm{eq}$ were contemporaneous indicates that Flare~7 may also be related to the increased number density of accelerated synchrotron particles.

Based on the discussion of the dominant source of jet energy in the flaring periods, we are able to investigate possible mechanisms generating the flares in the jet.
In Figure~\ref{fig: flares vs u_ratio}, we found that the dominant source of the energy of the jet changed while the flares occurred by comparing $u_{\rm p}$ to $u_{\rm B}$.
During the rise of Flare~3, $u_{\rm p}/u_{\rm B}$ decreased and the jet was converted from particle to magnetically dominated.
During the rise of Flare~7, the particle-energy-dominated jet reached equipartition, as $u_{\rm p}/u_{\rm B}$ changed from $>$$\sim$$10^5$ to $< 5.75$.
Note that most of the flares are generated when particle energy dominates in the jet.
While four flares (i.e., Flares~3, 4, 6, and 7) rise, the energy density of particles is $>$$\sim$$10^2$ to $>$$\sim$$10^5$ times larger than that of the magnetic field.
In such a large particle energy dominance environment in the jet, one expects that a particle acceleration mechanism (e.g., shock acceleration) may play a role \citep[e.g.,][]{Aller+1985}.
Based on the fact that the jet is dominated by particle energy in 15 epochs out of 17, the flares in the jet of CTA~102 can be generated by mainly particle acceleration mechanisms, e.g., shock acceleration.
It should be noted that the source was in equipartition condition or in the magnetic field energy dominance during the flaring periods.
Therefore, another candidate of particle acceleration during the flares is magnetic reconnection that dissipates the magnetic field and accelerates particles \citep[e.g.,][and references therein]{Blandford+2017}.


\section{Conclusions}    \label{sec:conclusions}

In this paper, we studied the physical origins of radio flares in the jet of the blazar CTA~102 based on the results from the multi-frequency observations at 2.6--343.5~GHz over the entire period 2012 November 20--2018 September 23 (MJD 56251--58384).
We estimated the magnetic field strengths and also investigated the relation between the magnetic field and the radio flare activity.

The 15~GHz data with densest sampling was used to investigate the variability properties of the source.
We found seven flares that occurred with the variability time-scale of $\sim$26--171~days in the 15~GHz light curve.
The variability time-scale of the flares imply that the emission regions of the source are compact, which enabled us to derive emitting region sizes in the range 0.06--0.19~mas, comparable to the VLBI core size.
We found that the flux densities of the VLBI core and those of the single-dish are correlated.
Our findings lead to the result that the variability of the observed emission may come from the compact core region.

The synchrotron spectrum of the source was investigated using quasi-simultaneous data.
We found that the synchrotron spectrum is self-absorbed and measured the turnover frequency as well as the peak flux density in the ranges $\sim$42--167~GHz and $\sim$0.9--10.2~Jy, respectively.
We derived synchrotron-self-absorption (SSA) magnetic field strengths of $9.20^{+9.17}_{-7.91}$~mG, $12.28^{+9.39}_{-11.25}$~mG, and $50.97^{+42.80}_{-33.92}$~mG for three epochs of 2013 December 24, 2014 February 28, and 2018 January 13, respectively.
We also derived 22 equipartition magnetic field strengths with the range $\sim$24--109~mG.

Based on the ratio of the magnetic field strengths, we investigated which source dominates the energy of the jet during the flaring periods.
We found that most equipartition magnetic field strengths are larger than the SSA magnetic field strengths, which indicates that particle energy mainly dominates in the SSA region.
Our results suggest that the flares in the jet of CTA~102 are mainly generated by particle acceleration.
We propose that the possible candidates of particle acceleration are shock acceleration and magnetic reconnection.

\section*{Acknowledgements}

We thank the anonymous referee for valuable comments and suggestions that helped to improve the paper.
We are grateful to the staff of the KVN who helped to operate the array and to correlate the data. The KVN is a facility operated by the KASI (Korea Astronomy and Space Science Institute). The KVN observations and correlations are supported through the high-speed network connections among the KVN sites provided by the KREONET (Korea Research Environment Open NETwork), which is managed and operated by the KISTI (Korea Institute of Science and Technology Information). 
This research has made use of data from the OVRO 40-m monitoring program which was supported in part by NASA grants NNX08AW31G, NNX11A043G and NNX14AQ89G, and NSF grants AST-0808050 and AST-1109911, and private funding from Caltech and the MPIfR. 
This publication makes use of data obtained at the Mets\"{a}hovi Radio Observatory, operated by Aalto University in Finland.  
I.A. acknowledges financial support from the Spanish "Ministerio de Ciencia e Innovaci\'{o}n” (MCINN) through the “Center of Excellence Severo Ochoa” award for the Instituto de Astrof\'{i}sica de Andaluc\'{i}a-CSIC (SEV-2017-0709). Acquisition and reduction of the MAPCAT data was supported in part by MICINN through grants AYA2016-80889-P and PID2019-107847RB-C44. The POLAMI observations were carried out at the IRAM 30m Telescope. IRAM is supported by INSU/CNRS (France), MPG (Germany) and IGN (Spain). 
The Submillimeter Array is a joint project between the Smithsonian Astrophysical Observatory and the Academia Sinica Institute of Astronomy and Astrophysics and is funded by the Smithsonian Institution and the Academia Sinica. 
This paper makes use of the following ALMA data: ADS/JAO.ALMA\#2011.0.00001.CAL. ALMA is a partnership of ESO (representing its member states), NSF (USA) and NINS (Japan), together with NRC (Canada), MOST and ASIAA (Taiwan), and KASI (Republic of Korea), in cooperation with the Republic of Chile. The Joint ALMA Observatory is operated by ESO, AUI/NRAO and NAOJ.  
This research has made use of data from the MOJAVE database that is maintained by the MOJAVE team \citep{Lister+2018}.    
This study makes use of VLBA data from the VLBA-BU Blazar Monitoring Program (BEAM-ME and VLBA-BU-BLAZAR;
\url{http://www.bu.edu/blazars/BEAM-ME.html}), funded by NASA through the Fermi Guest Investigator Program. The VLBA is an instrument of the National Radio Astronomy Observatory. The National Radio Astronomy Observatory is a facility of the National Science Foundation operated by Associated Universities, Inc.  
This work was supported by the National Research Foundation of Korea (NRF) grant funded by the Korea government (MIST) (2020R1A2C2009003).

\section*{Data Availability}


The data underlying this article will be shared on reasonable request to the corresponding author.
And the SMA data are available in Submillimeter Calibrator List at \url{http://sma1.sma.hawaii.edu/callist/callist.html}; for questions regarding their availability, please contact Mark Gurwell (mgurwell@cfa.harvard.edu).
The ALMA data are available in ALMA Calibrator Source Catalogue at \url{https://almascience.eso.org/sc/}.
The VLBA 15~GHz data collected by the MOJAVE program are available at \url{https://www.physics.purdue.edu/MOJAVE/}.
The VLBA 43~GHz data collected by the VLBA-BU-BLAZAR monitoring program are available at \url{https://www.bu.edu/blazars/VLBAproject.html}.


\bibliographystyle{mnras}
\bibliography{refs} 

\begin{thebibliography}{}
\makeatletter
\relax
\def\mn@urlcharsother{\let\do\@makeother \do\$\do\&\do\#\do\^\do\_\do\%\do\~}
\def\mn@doi{\begingroup\mn@urlcharsother \@ifnextchar [ {\mn@doi@}
  {\mn@doi@[]}}
\def\mn@doi@[#1]#2{\def\@tempa{#1}\ifx\@tempa\@empty \href
  {http://dx.doi.org/#2} {doi:#2}\else \href {http://dx.doi.org/#2} {#1}\fi
  \endgroup}
\def\mn@eprint#1#2{\mn@eprint@#1:#2::\@nil}
\def\mn@eprint@arXiv#1{\href {http://arxiv.org/abs/#1} {{\tt arXiv:#1}}}
\def\mn@eprint@dblp#1{\href {http://dblp.uni-trier.de/rec/bibtex/#1.xml}
  {dblp:#1}}
\def\mn@eprint@#1:#2:#3:#4\@nil{\def\@tempa {#1}\def\@tempb {#2}\def\@tempc
  {#3}\ifx \@tempc \@empty \let \@tempc \@tempb \let \@tempb \@tempa \fi \ifx
  \@tempb \@empty \def\@tempb {arXiv}\fi \@ifundefined
  {mn@eprint@\@tempb}{\@tempb:\@tempc}{\expandafter \expandafter \csname
  mn@eprint@\@tempb\endcsname \expandafter{\@tempc}}}

\bibitem[\protect\citeauthoryear{{Agudo}, {Thum}, {Wiesemeyer}  \&
  {Krichbaum}}{{Agudo} et~al.}{2010}]{Agudo+2010}
{Agudo} I.,  {Thum} C.,  {Wiesemeyer} H.,   {Krichbaum} T.~P.,  2010, \mn@doi
  [\apjs] {10.1088/0067-0049/189/1/1}, \href
  {https://ui.adsabs.harvard.edu/abs/2010ApJS..189....1A} {189, 1}

\bibitem[\protect\citeauthoryear{{Agudo}, {Thum}, {G{\'o}mez}  \&
  {Wiesemeyer}}{{Agudo} et~al.}{2014}]{Agudo+2014}
{Agudo} I.,  {Thum} C.,  {G{\'o}mez} J.~L.,   {Wiesemeyer} H.,  2014, \mn@doi
  [\aap] {10.1051/0004-6361/201423366}, \href
  {https://ui.adsabs.harvard.edu/abs/2014A&A...566A..59A} {566, A59}

\bibitem[\protect\citeauthoryear{{Agudo}, {Thum}, {Ramakrishnan}, {Molina},
  {Casadio}  \& {G{\'o}mez}}{{Agudo} et~al.}{2018a}]{Agudo+2018a}
{Agudo} I.,  {Thum} C.,  {Ramakrishnan} V.,  {Molina} S.~N.,  {Casadio} C.,
  {G{\'o}mez} J.~L.,  2018a, \mn@doi [\mnras] {10.1093/mnras/stx2437}, \href
  {https://ui.adsabs.harvard.edu/abs/2018MNRAS.473.1850A} {473, 1850}

\bibitem[\protect\citeauthoryear{{Agudo} et~al.,}{{Agudo}
  et~al.}{2018b}]{Agudo+2018b}
{Agudo} I.,  et~al., 2018b, \mn@doi [\mnras] {10.1093/mnras/stx2435}, \href
  {https://ui.adsabs.harvard.edu/abs/2018MNRAS.474.1427A} {474, 1427}

\bibitem[\protect\citeauthoryear{{Algaba}, {Gabuzda}  \& {Smith}}{{Algaba}
  et~al.}{2012}]{Algaba+2012}
{Algaba} J.~C.,  {Gabuzda} D.~C.,   {Smith} P.~S.,  2012, \mn@doi [\mnras]
  {10.1111/j.1365-2966.2011.20061.x}, \href
  {https://ui.adsabs.harvard.edu/abs/2012MNRAS.420..542A} {420, 542}

\bibitem[\protect\citeauthoryear{{Algaba} et~al.,}{{Algaba}
  et~al.}{2018a}]{Algaba+2018a}
{Algaba} J.-C.,  et~al., 2018a, \mn@doi [\apj] {10.3847/1538-4357/aa9e50},
  \href {https://ui.adsabs.harvard.edu/abs/2018ApJ...852...30A} {852, 30}

\bibitem[\protect\citeauthoryear{{Algaba} et~al.,}{{Algaba}
  et~al.}{2018b}]{Algaba+2018b}
{Algaba} J.-C.,  et~al., 2018b, \mn@doi [\apj] {10.3847/1538-4357/aac2e7},
  \href {https://ui.adsabs.harvard.edu/abs/2018ApJ...859..128A} {859, 128}

\bibitem[\protect\citeauthoryear{{Aller}, {Aller}  \& {Hughes}}{{Aller}
  et~al.}{1985}]{Aller+1985}
{Aller} H.~D.,  {Aller} M.~F.,   {Hughes} P.~A.,  1985, \mn@doi [\apj]
  {10.1086/163610}, \href
  {https://ui.adsabs.harvard.edu/abs/1985ApJ...298..296A} {298, 296}

\bibitem[\protect\citeauthoryear{{Angelakis} et~al.,}{{Angelakis}
  et~al.}{2019}]{Angelakis+2019}
{Angelakis} E.,  et~al., 2019, \mn@doi [\aap] {10.1051/0004-6361/201834363},
  \href {https://ui.adsabs.harvard.edu/abs/2019A&A...626A..60A} {626, A60}

\bibitem[\protect\citeauthoryear{{Blandford} \& {K{\"o}nigl}}{{Blandford} \&
  {K{\"o}nigl}}{1979}]{Blandford&Konigl+1979}
{Blandford} R.~D.,  {K{\"o}nigl} A.,  1979, \mn@doi [\apj] {10.1086/157262},
  \href {https://ui.adsabs.harvard.edu/abs/1979ApJ...232...34B} {232, 34}

\bibitem[\protect\citeauthoryear{{Blandford} \& {Payne}}{{Blandford} \&
  {Payne}}{1982}]{Blandford&Payne+1982}
{Blandford} R.~D.,  {Payne} D.~G.,  1982, \mn@doi [\mnras]
  {10.1093/mnras/199.4.883}, \href
  {https://ui.adsabs.harvard.edu/abs/1982MNRAS.199..883B} {199, 883}

\bibitem[\protect\citeauthoryear{{Blandford} \& {Znajek}}{{Blandford} \&
  {Znajek}}{1977}]{Blandford&Znajek+1977}
{Blandford} R.~D.,  {Znajek} R.~L.,  1977, \mn@doi [\mnras]
  {10.1093/mnras/179.3.433}, \href
  {https://ui.adsabs.harvard.edu/abs/1977MNRAS.179..433B} {179, 433}

\bibitem[\protect\citeauthoryear{{Blandford}, {Yuan}, {Hoshino}  \&
  {Sironi}}{{Blandford} et~al.}{2017}]{Blandford+2017}
{Blandford} R.,  {Yuan} Y.,  {Hoshino} M.,   {Sironi} L.,  2017, \mn@doi [\ssr]
  {10.1007/s11214-017-0376-2}, \href
  {https://ui.adsabs.harvard.edu/abs/2017SSRv..207..291B} {207, 291}

\bibitem[\protect\citeauthoryear{{Blandford}, {Meier}  \&
  {Readhead}}{{Blandford} et~al.}{2019}]{Blandford+2019}
{Blandford} R.,  {Meier} D.,   {Readhead} A.,  2019, \mn@doi [\araa]
  {10.1146/annurev-astro-081817-051948}, \href
  {https://ui.adsabs.harvard.edu/abs/2019ARA&A..57..467B} {57, 467}

\bibitem[\protect\citeauthoryear{{Boettcher}, {Harris}  \&
  {Krawczynski}}{{Boettcher} et~al.}{2012}]{Boettcher+2012}
{Boettcher} M.,  {Harris} D.~E.,   {Krawczynski} H.,  2012, {Relativistic Jets
  from Active Galactic Nuclei}

\bibitem[\protect\citeauthoryear{{Casadio} et~al.,}{{Casadio}
  et~al.}{2015}]{Casadio+2015}
{Casadio} C.,  et~al., 2015, \mn@doi [\apj] {10.1088/0004-637X/813/1/51}, \href
  {https://ui.adsabs.harvard.edu/abs/2015ApJ...813...51C} {813, 51}

\bibitem[\protect\citeauthoryear{{Casadio} et~al.,}{{Casadio}
  et~al.}{2019}]{Casadio+2019}
{Casadio} C.,  et~al., 2019, \mn@doi [\aap] {10.1051/0004-6361/201834519},
  \href {https://ui.adsabs.harvard.edu/abs/2019A&A...622A.158C} {622, A158}

\bibitem[\protect\citeauthoryear{{Celotti} \& {Fabian}}{{Celotti} \&
  {Fabian}}{1993}]{Celotti&Fabian+1993}
{Celotti} A.,  {Fabian} A.~C.,  1993, \mn@doi [\mnras]
  {10.1093/mnras/264.1.228}, \href
  {https://ui.adsabs.harvard.edu/abs/1993MNRAS.264..228C} {264, 228}

\bibitem[\protect\citeauthoryear{{Celotti} \& {Ghisellini}}{{Celotti} \&
  {Ghisellini}}{2008}]{Celotti&Ghisellini+2008}
{Celotti} A.,  {Ghisellini} G.,  2008, \mn@doi [\mnras]
  {10.1111/j.1365-2966.2007.12758.x}, \href
  {https://ui.adsabs.harvard.edu/abs/2008MNRAS.385..283C} {385, 283}

\bibitem[\protect\citeauthoryear{{Condon} \& {Ransom}}{{Condon} \&
  {Ransom}}{2016}]{Condon&Ransom+2016}
{Condon} J.~J.,  {Ransom} S.~M.,  2016, {Essential Radio Astronomy}

\bibitem[\protect\citeauthoryear{{D'Ammando} et~al.,}{{D'Ammando}
  et~al.}{2019}]{D'Ammando+2019}
{D'Ammando} F.,  et~al., 2019, \mn@doi [\mnras] {10.1093/mnras/stz2792}, \href
  {https://ui.adsabs.harvard.edu/abs/2019MNRAS.490.5300D} {490, 5300}

\bibitem[\protect\citeauthoryear{{Foreman-Mackey}, {Hogg}, {Lang}  \&
  {Goodman}}{{Foreman-Mackey} et~al.}{2013}]{Foreman-Mackey+2013}
{Foreman-Mackey} D.,  {Hogg} D.~W.,  {Lang} D.,   {Goodman} J.,  2013, \mn@doi
  [\pasp] {10.1086/670067}, \href
  {https://ui.adsabs.harvard.edu/abs/2013PASP..125..306F} {125, 306}

\bibitem[\protect\citeauthoryear{{Fromm} et~al.,}{{Fromm}
  et~al.}{2011}]{Fromm+2011}
{Fromm} C.~M.,  et~al., 2011, \mn@doi [\aap] {10.1051/0004-6361/201116857},
  \href {https://ui.adsabs.harvard.edu/abs/2011A&A...531A..95F} {531, A95}

\bibitem[\protect\citeauthoryear{{Fromm} et~al.,}{{Fromm}
  et~al.}{2013a}]{Fromm+2013a}
{Fromm} C.~M.,  et~al., 2013a, \mn@doi [\aap] {10.1051/0004-6361/201219913},
  \href {https://ui.adsabs.harvard.edu/abs/2013A&A...551A..32F} {551, A32}

\bibitem[\protect\citeauthoryear{{Fromm}, {Ros}, {Perucho}, {Savolainen},
  {Mimica}, {Kadler}, {Lobanov}  \& {Zensus}}{{Fromm}
  et~al.}{2013b}]{Fromm+2013b}
{Fromm} C.~M.,  {Ros} E.,  {Perucho} M.,  {Savolainen} T.,  {Mimica} P.,
  {Kadler} M.,  {Lobanov} A.~P.,   {Zensus} J.~A.,  2013b, \mn@doi [\aap]
  {10.1051/0004-6361/201321784}, \href
  {https://ui.adsabs.harvard.edu/abs/2013A&A...557A.105F} {557, A105}

\bibitem[\protect\citeauthoryear{{Fuhrmann} et~al.,}{{Fuhrmann}
  et~al.}{2016}]{Fuhrmann+2016}
{Fuhrmann} L.,  et~al., 2016, \mn@doi [\aap] {10.1051/0004-6361/201528034},
  \href {https://ui.adsabs.harvard.edu/abs/2016A&A...596A..45F} {596, A45}

\bibitem[\protect\citeauthoryear{{Gasparyan}, {Sahakyan}, {Baghmanyan}  \&
  {Zargaryan}}{{Gasparyan} et~al.}{2018}]{Gasparyan+2018}
{Gasparyan} S.,  {Sahakyan} N.,  {Baghmanyan} V.,   {Zargaryan} D.,  2018,
  \mn@doi [\apj] {10.3847/1538-4357/aad234}, \href
  {https://ui.adsabs.harvard.edu/abs/2018ApJ...863..114G} {863, 114}

\bibitem[\protect\citeauthoryear{{Ghisellini}}{{Ghisellini}}{2013}]{Ghisellini+2013}
{Ghisellini} G.,  2013, {Radiative Processes in High Energy Astrophysics}.
 Vol. 873, \mn@doi{10.1007/978-3-319-00612-3, }

\bibitem[\protect\citeauthoryear{{Gurwell}, {Peck}, {Hostler}, {Darrah}  \&
  {Katz}}{{Gurwell} et~al.}{2007}]{Gurwell+2007}
{Gurwell} M.~A.,  {Peck} A.~B.,  {Hostler} S.~R.,  {Darrah} M.~R.,   {Katz}
  C.~A.,  2007, in {Baker} A.~J.,  {Glenn} J.,  {Harris} A.~I.,  {Mangum}
  J.~G.,   {Yun} M.~S.,  eds,  Astronomical Society of the Pacific Conference
  Series Vol. 375, From Z-Machines to ALMA: (Sub)Millimeter Spectroscopy of
  Galaxies. p.~234

\bibitem[\protect\citeauthoryear{{Hirotani}}{{Hirotani}}{2005}]{Hirotani+2005}
{Hirotani} K.,  2005, \mn@doi [\apj] {10.1086/426497}, \href
  {https://ui.adsabs.harvard.edu/abs/2005ApJ...619...73H} {619, 73}

\bibitem[\protect\citeauthoryear{{Hodgson} et~al.,}{{Hodgson}
  et~al.}{2017}]{Hodgson+2017}
{Hodgson} J.~A.,  et~al., 2017, \mn@doi [\aap] {10.1051/0004-6361/201526727},
  \href {https://ui.adsabs.harvard.edu/abs/2017A&A...597A..80H} {597, A80}

\bibitem[\protect\citeauthoryear{{Homan} et~al.,}{{Homan}
  et~al.}{2006}]{Homan+2006}
{Homan} D.~C.,  et~al., 2006, \mn@doi [\apjl] {10.1086/504715}, \href
  {https://ui.adsabs.harvard.edu/abs/2006ApJ...642L.115H} {642, L115}

\bibitem[\protect\citeauthoryear{{Hovatta} \& {Lindfors}}{{Hovatta} \&
  {Lindfors}}{2019}]{Hovatta+2019}
{Hovatta} T.,  {Lindfors} E.,  2019, \mn@doi [\nar]
  {10.1016/j.newar.2020.101541}, \href
  {https://ui.adsabs.harvard.edu/abs/2019NewAR..8701541H} {87, 101541}

\bibitem[\protect\citeauthoryear{{Hovatta}, {Nieppola}, {Tornikoski},
  {Valtaoja}, {Aller}  \& {Aller}}{{Hovatta} et~al.}{2008}]{Hovatta+2008}
{Hovatta} T.,  {Nieppola} E.,  {Tornikoski} M.,  {Valtaoja} E.,  {Aller} M.~F.,
    {Aller} H.~D.,  2008, \mn@doi [\aap] {10.1051/0004-6361:200809806}, \href
  {https://ui.adsabs.harvard.edu/abs/2008A&A...485...51H} {485, 51}

\bibitem[\protect\citeauthoryear{{Hovatta}, {Valtaoja}, {Tornikoski}  \&
  {L{\"a}hteenm{\"a}ki}}{{Hovatta} et~al.}{2009}]{Hovatta+2009}
{Hovatta} T.,  {Valtaoja} E.,  {Tornikoski} M.,   {L{\"a}hteenm{\"a}ki} A.,
  2009, \mn@doi [\aap] {10.1051/0004-6361:200811150}, \href
  {https://ui.adsabs.harvard.edu/abs/2009A&A...494..527H} {494, 527}

\bibitem[\protect\citeauthoryear{{Hovatta}, {Lister}, {Aller}, {Aller},
  {Homan}, {Kovalev}, {Pushkarev}  \& {Savolainen}}{{Hovatta}
  et~al.}{2012}]{Hovatta+2012}
{Hovatta} T.,  {Lister} M.~L.,  {Aller} M.~F.,  {Aller} H.~D.,  {Homan} D.~C.,
  {Kovalev} Y.~Y.,  {Pushkarev} A.~B.,   {Savolainen} T.,  2012, \mn@doi [\aj]
  {10.1088/0004-6256/144/4/105}, \href
  {https://ui.adsabs.harvard.edu/abs/2012AJ....144..105H} {144, 105}

\bibitem[\protect\citeauthoryear{{Jorstad} et~al.,}{{Jorstad}
  et~al.}{2005}]{Jorstad+2005}
{Jorstad} S.~G.,  et~al., 2005, \mn@doi [\aj] {10.1086/444593}, \href
  {https://ui.adsabs.harvard.edu/abs/2005AJ....130.1418J} {130, 1418}

\bibitem[\protect\citeauthoryear{{Jorstad} et~al.,}{{Jorstad}
  et~al.}{2017}]{Jorstad+2017}
{Jorstad} S.~G.,  et~al., 2017, \mn@doi [\apj] {10.3847/1538-4357/aa8407},
  \href {https://ui.adsabs.harvard.edu/abs/2017ApJ...846...98J} {846, 98}

\bibitem[\protect\citeauthoryear{{Kang}, {Chen}  \& {Wu}}{{Kang}
  et~al.}{2014}]{Kang+2014}
{Kang} S.-J.,  {Chen} L.,   {Wu} Q.,  2014, \mn@doi [\apjs]
  {10.1088/0067-0049/215/1/5}, \href
  {https://ui.adsabs.harvard.edu/abs/2014ApJS..215....5K} {215, 5}

\bibitem[\protect\citeauthoryear{{Kang} et~al.,}{{Kang}
  et~al.}{2021}]{Kang+2021}
{Kang} S.,  et~al., 2021, \mn@doi [\aap] {10.1051/0004-6361/202040198}, \href
  {https://ui.adsabs.harvard.edu/abs/2021A&A...651A..74K} {651, A74}

\bibitem[\protect\citeauthoryear{{Kataoka} \& {Stawarz}}{{Kataoka} \&
  {Stawarz}}{2005}]{Kataoka&Stawarz+2005}
{Kataoka} J.,  {Stawarz} {\L}.,  2005, \mn@doi [\apj] {10.1086/428083}, \href
  {https://ui.adsabs.harvard.edu/abs/2005ApJ...622..797K} {622, 797}

\bibitem[\protect\citeauthoryear{{Kellermann} \& {Pauliny-Toth}}{{Kellermann}
  \& {Pauliny-Toth}}{1969}]{Kellermann&Pauliny-Toth+1969}
{Kellermann} K.~I.,  {Pauliny-Toth} I.~I.~K.,  1969, \mn@doi [\apjl]
  {10.1086/180305}, \href
  {https://ui.adsabs.harvard.edu/abs/1969ApJ...155L..71K} {155, L71}

\bibitem[\protect\citeauthoryear{{Kravchenko} et~al.,}{{Kravchenko}
  et~al.}{2020}]{Kravchenko+2020}
{Kravchenko} E.~V.,  et~al., 2020, \mn@doi [\apj] {10.3847/1538-4357/ab7dae},
  \href {https://ui.adsabs.harvard.edu/abs/2020ApJ...893...68K} {893, 68}

\bibitem[\protect\citeauthoryear{{L{\"a}hteenm{\"a}ki} \&
  {Valtaoja}}{{L{\"a}hteenm{\"a}ki} \&
  {Valtaoja}}{1999}]{Lahteenmaki&Valtaoja+1999}
{L{\"a}hteenm{\"a}ki} A.,  {Valtaoja} E.,  1999, \mn@doi [\apj]
  {10.1086/307587}, \href
  {https://ui.adsabs.harvard.edu/abs/1999ApJ...521..493L} {521, 493}

\bibitem[\protect\citeauthoryear{{L{\"a}hteenm{\"a}ki}, {Valtaoja}  \&
  {Wiik}}{{L{\"a}hteenm{\"a}ki} et~al.}{1999}]{Lahteenmaki+1999}
{L{\"a}hteenm{\"a}ki} A.,  {Valtaoja} E.,   {Wiik} K.,  1999, \mn@doi [\apj]
  {10.1086/306649}, \href
  {https://ui.adsabs.harvard.edu/abs/1999ApJ...511..112L} {511, 112}

\bibitem[\protect\citeauthoryear{{Lee} et~al.,}{{Lee}
  et~al.}{2011}]{LeeSS+2011}
{Lee} S.-S.,  et~al., 2011, \mn@doi [\pasp] {10.1086/663326}, \href
  {https://ui.adsabs.harvard.edu/abs/2011PASP..123.1398L} {123, 1398}

\bibitem[\protect\citeauthoryear{{Lee} et~al.,}{{Lee}
  et~al.}{2016a}]{LeeSS+2016b}
{Lee} S.-S.,  et~al., 2016a, \mn@doi [\apjs] {10.3847/0067-0049/227/1/8}, \href
  {https://ui.adsabs.harvard.edu/abs/2016ApJS..227....8L} {227, 8}

\bibitem[\protect\citeauthoryear{{Lee}, {Lobanov}, {Krichbaum}  \&
  {Zensus}}{{Lee} et~al.}{2016b}]{LeeSS+2016a}
{Lee} S.-S.,  {Lobanov} A.~P.,  {Krichbaum} T.~P.,   {Zensus} J.~A.,  2016b,
  \mn@doi [\apj] {10.3847/0004-637X/826/2/135}, \href
  {https://ui.adsabs.harvard.edu/abs/2016ApJ...826..135L} {826, 135}

\bibitem[\protect\citeauthoryear{{Lee}, {Sohn}, {Jung}, {Byun}  \& {Lee}}{{Lee}
  et~al.}{2017a}]{LeeJA+2017}
{Lee} J.~A.,  {Sohn} B.~W.,  {Jung} T.,  {Byun} D.-Y.,   {Lee} J.~W.,  2017a,
  \mn@doi [\apjs] {10.3847/1538-4365/228/2/22}, \href
  {https://ui.adsabs.harvard.edu/abs/2017ApJS..228...22L} {228, 22}

\bibitem[\protect\citeauthoryear{{Lee}, {Lee}, {Hodgson}, {Kim}, {Algaba},
  {Kang}, {Kang}  \& {Kim}}{{Lee} et~al.}{2017b}]{LeeJW+2017}
{Lee} J.~W.,  {Lee} S.-S.,  {Hodgson} J.~A.,  {Kim} D.-W.,  {Algaba} J.-C.,
  {Kang} S.,  {Kang} J.,   {Kim} S.~S.,  2017b, \mn@doi [\apj]
  {10.3847/1538-4357/aa72f7}, \href
  {https://ui.adsabs.harvard.edu/abs/2017ApJ...841..119L} {841, 119}

\bibitem[\protect\citeauthoryear{{Lee} et~al.,}{{Lee}
  et~al.}{2020}]{LeeJW+2020}
{Lee} J.~W.,  et~al., 2020, \mn@doi [\apj] {10.3847/1538-4357/abb4e5}, \href
  {https://ui.adsabs.harvard.edu/abs/2020ApJ...902..104L} {902, 104}

\bibitem[\protect\citeauthoryear{{Li} et~al.,}{{Li} et~al.}{2018}]{Li+2018}
{Li} X.,  et~al., 2018, \mn@doi [\apj] {10.3847/1538-4357/aaa5ac}, \href
  {https://ui.adsabs.harvard.edu/abs/2018ApJ...854...17L} {854, 17}

\bibitem[\protect\citeauthoryear{{Liodakis} et~al.,}{{Liodakis}
  et~al.}{2017}]{Liodakis+2017}
{Liodakis} I.,  et~al., 2017, \mn@doi [\mnras] {10.1093/mnras/stx002}, \href
  {https://ui.adsabs.harvard.edu/abs/2017MNRAS.466.4625L} {466, 4625}

\bibitem[\protect\citeauthoryear{{Liodakis}, {Hovatta}, {Huppenkothen},
  {Kiehlmann}, {Max-Moerbeck}  \& {Readhead}}{{Liodakis}
  et~al.}{2018}]{Liodakis+2018}
{Liodakis} I.,  {Hovatta} T.,  {Huppenkothen} D.,  {Kiehlmann} S.,
  {Max-Moerbeck} W.,   {Readhead} A. C.~S.,  2018, \mn@doi [\apj]
  {10.3847/1538-4357/aae2b7}, \href
  {https://ui.adsabs.harvard.edu/abs/2018ApJ...866..137L} {866, 137}

\bibitem[\protect\citeauthoryear{{Lister} \& {Homan}}{{Lister} \&
  {Homan}}{2005}]{Lister+2005}
{Lister} M.~L.,  {Homan} D.~C.,  2005, \mn@doi [\aj] {10.1086/432969}, \href
  {https://ui.adsabs.harvard.edu/abs/2005AJ....130.1389L} {130, 1389}

\bibitem[\protect\citeauthoryear{{Lister} et~al.,}{{Lister}
  et~al.}{2009a}]{Lister+2009a}
{Lister} M.~L.,  et~al., 2009a, \mn@doi [\aj] {10.1088/0004-6256/137/3/3718},
  \href {https://ui.adsabs.harvard.edu/abs/2009AJ....137.3718L} {137, 3718}

\bibitem[\protect\citeauthoryear{{Lister} et~al.,}{{Lister}
  et~al.}{2009b}]{Lister+2009b}
{Lister} M.~L.,  et~al., 2009b, \mn@doi [\aj] {10.1088/0004-6256/138/6/1874},
  \href {https://ui.adsabs.harvard.edu/abs/2009AJ....138.1874L} {138, 1874}

\bibitem[\protect\citeauthoryear{{Lister} et~al.,}{{Lister}
  et~al.}{2016}]{Lister+2016}
{Lister} M.~L.,  et~al., 2016, \mn@doi [\aj] {10.3847/0004-6256/152/1/12},
  \href {https://ui.adsabs.harvard.edu/abs/2016AJ....152...12L} {152, 12}

\bibitem[\protect\citeauthoryear{{Lister}, {Aller}, {Aller}, {Hodge}, {Homan},
  {Kovalev}, {Pushkarev}  \& {Savolainen}}{{Lister} et~al.}{2018}]{Lister+2018}
{Lister} M.~L.,  {Aller} M.~F.,  {Aller} H.~D.,  {Hodge} M.~A.,  {Homan} D.~C.,
   {Kovalev} Y.~Y.,  {Pushkarev} A.~B.,   {Savolainen} T.,  2018, \mn@doi
  [\apjs] {10.3847/1538-4365/aa9c44}, \href
  {https://ui.adsabs.harvard.edu/abs/2018ApJS..234...12L} {234, 12}

\bibitem[\protect\citeauthoryear{{Lobanov}}{{Lobanov}}{1998}]{Lobanov+1998}
{Lobanov} A.~P.,  1998, \aap, \href
  {https://ui.adsabs.harvard.edu/abs/1998A&A...330...79L} {330, 79}

\bibitem[\protect\citeauthoryear{{Longair}}{{Longair}}{1994}]{Longair+1994}
{Longair} M.~S.,  1994, {High energy astrophysics. Volume 2. Stars, the Galaxy
  and the interstellar medium.}.
 Vol. 2

\bibitem[\protect\citeauthoryear{{Marscher}}{{Marscher}}{1983}]{Marscher+1983}
{Marscher} A.~P.,  1983, \mn@doi [\apj] {10.1086/160597}, \href
  {https://ui.adsabs.harvard.edu/abs/1983ApJ...264..296M} {264, 296}

\bibitem[\protect\citeauthoryear{{Marscher} \& {Gear}}{{Marscher} \&
  {Gear}}{1985}]{Marscher&Gear+1985}
{Marscher} A.~P.,  {Gear} W.~K.,  1985, \mn@doi [\apj] {10.1086/163592}, \href
  {https://ui.adsabs.harvard.edu/abs/1985ApJ...298..114M} {298, 114}

\bibitem[\protect\citeauthoryear{{Marscher} et~al.,}{{Marscher}
  et~al.}{2008}]{Marscher+2008}
{Marscher} A.~P.,  et~al., 2008, \mn@doi [\nat] {10.1038/nature06895}, \href
  {https://ui.adsabs.harvard.edu/abs/2008Natur.452..966M} {452, 966}

\bibitem[\protect\citeauthoryear{{Mastichiadis} \& {Kirk}}{{Mastichiadis} \&
  {Kirk}}{1997}]{Mastichiadis&Kirk+1997}
{Mastichiadis} A.,  {Kirk} J.~G.,  1997, \aap, \href
  {https://ui.adsabs.harvard.edu/abs/1997A&A...320...19M} {320, 19}

\bibitem[\protect\citeauthoryear{{O'Sullivan} \& {Gabuzda}}{{O'Sullivan} \&
  {Gabuzda}}{2009}]{O'Sullivan&Gabuzda+2009}
{O'Sullivan} S.~P.,  {Gabuzda} D.~C.,  2009, \mn@doi [\mnras]
  {10.1111/j.1365-2966.2009.15428.x}, \href
  {https://ui.adsabs.harvard.edu/abs/2009MNRAS.400...26O} {400, 26}

\bibitem[\protect\citeauthoryear{{Pacholczyk}}{{Pacholczyk}}{1970}]{Pacholczyk+1970}
{Pacholczyk} A.~G.,  1970, {Radio astrophysics. Nonthermal processes in
  galactic and extragalactic sources}

\bibitem[\protect\citeauthoryear{{Park} et~al.,}{{Park}
  et~al.}{2018}]{Park+2018}
{Park} J.,  et~al., 2018, \mn@doi [\apj] {10.3847/1538-4357/aac490}, \href
  {https://ui.adsabs.harvard.edu/abs/2018ApJ...860..112P} {860, 112}

\bibitem[\protect\citeauthoryear{{Prince}, {Raman}, {Hahn}, {Gupta}  \&
  {Majumdar}}{{Prince} et~al.}{2018}]{Prince+2018}
{Prince} R.,  {Raman} G.,  {Hahn} J.,  {Gupta} N.,   {Majumdar} P.,  2018,
  \mn@doi [\apj] {10.3847/1538-4357/aadadb}, \href
  {https://ui.adsabs.harvard.edu/abs/2018ApJ...866...16P} {866, 16}

\bibitem[\protect\citeauthoryear{{Pushkarev}, {Hovatta}, {Kovalev}, {Lister},
  {Lobanov}, {Savolainen}  \& {Zensus}}{{Pushkarev}
  et~al.}{2012}]{Pushkarev+2012}
{Pushkarev} A.~B.,  {Hovatta} T.,  {Kovalev} Y.~Y.,  {Lister} M.~L.,  {Lobanov}
  A.~P.,  {Savolainen} T.,   {Zensus} J.~A.,  2012, \mn@doi [\aap]
  {10.1051/0004-6361/201219173}, \href
  {https://ui.adsabs.harvard.edu/abs/2012A&A...545A.113P} {545, A113}

\bibitem[\protect\citeauthoryear{{Raiteri} et~al.,}{{Raiteri}
  et~al.}{2017}]{Raiteri+2017}
{Raiteri} C.~M.,  et~al., 2017, \mn@doi [\nat] {10.1038/nature24623}, \href
  {https://ui.adsabs.harvard.edu/abs/2017Natur.552..374R} {552, 374}

\bibitem[\protect\citeauthoryear{{Rani} et~al.,}{{Rani}
  et~al.}{2013}]{Rani+2013}
{Rani} B.,  et~al., 2013, \mn@doi [\aap] {10.1051/0004-6361/201321058}, \href
  {https://ui.adsabs.harvard.edu/abs/2013A&A...552A..11R} {552, A11}

\bibitem[\protect\citeauthoryear{{Rantakyr{\"o}}, {Wiik}, {Tornikoski},
  {Valtaoja}  \& {B{\r{a}}{\r{a}}th}}{{Rantakyr{\"o}}
  et~al.}{2003}]{Rantakyro+2003}
{Rantakyr{\"o}} F.~T.,  {Wiik} K.,  {Tornikoski} M.,  {Valtaoja} E.,
  {B{\r{a}}{\r{a}}th} L.~B.,  2003, \mn@doi [\aap]
  {10.1051/0004-6361:20030579}, \href
  {https://ui.adsabs.harvard.edu/abs/2003A&A...405..473R} {405, 473}

\bibitem[\protect\citeauthoryear{{Readhead}}{{Readhead}}{1994}]{Readhead+1994}
{Readhead} A. C.~S.,  1994, \mn@doi [\apj] {10.1086/174038}, \href
  {https://ui.adsabs.harvard.edu/abs/1994ApJ...426...51R} {426, 51}

\bibitem[\protect\citeauthoryear{{Richards} et~al.,}{{Richards}
  et~al.}{2011}]{Richards+2011}
{Richards} J.~L.,  et~al., 2011, \mn@doi [\apjs] {10.1088/0067-0049/194/2/29},
  \href {https://ui.adsabs.harvard.edu/abs/2011ApJS..194...29R} {194, 29}

\bibitem[\protect\citeauthoryear{{Schmidt}}{{Schmidt}}{1965}]{Schmidt+1965}
{Schmidt} M.,  1965, \mn@doi [\apj] {10.1086/148217}, \href
  {https://ui.adsabs.harvard.edu/abs/1965ApJ...141.1295S} {141, 1295}

\bibitem[\protect\citeauthoryear{{Shepherd}}{{Shepherd}}{1997}]{Shepherd+1997}
{Shepherd} M.~C.,  1997, in {Hunt} G.,  {Payne} H.,  eds,  Astronomical Society
  of the Pacific Conference Series Vol. 125, Astronomical Data Analysis
  Software and Systems VI. p.~77

\bibitem[\protect\citeauthoryear{{Singal}}{{Singal}}{2009}]{Singal+2009}
{Singal} A.~K.,  2009, \mn@doi [\apjl] {10.1088/0004-637X/703/2/L109}, \href
  {https://ui.adsabs.harvard.edu/abs/2009ApJ...703L.109S} {703, L109}

\bibitem[\protect\citeauthoryear{{Tavecchio} \& {Ghisellini}}{{Tavecchio} \&
  {Ghisellini}}{2014}]{Tavecchio&Ghisellini+2014}
{Tavecchio} F.,  {Ghisellini} G.,  2014, \mn@doi [\mnras]
  {10.1093/mnras/stu1196}, \href
  {https://ui.adsabs.harvard.edu/abs/2014MNRAS.443.1224T} {443, 1224}

\bibitem[\protect\citeauthoryear{{Teraesranta} et~al.,}{{Teraesranta}
  et~al.}{1998}]{Teraesranta+1998}
{Teraesranta} H.,  et~al., 1998, \mn@doi [\aaps] {10.1051/aas:1998297}, \href
  {https://ui.adsabs.harvard.edu/abs/1998A&AS..132..305T} {132, 305}

\bibitem[\protect\citeauthoryear{{Thum}, {Wiesemeyer}, {Paubert}, {Navarro}  \&
  {Morris}}{{Thum} et~al.}{2008}]{Thum+2008}
{Thum} C.,  {Wiesemeyer} H.,  {Paubert} G.,  {Navarro} S.,   {Morris} D.,
  2008, \mn@doi [\pasp] {10.1086/590190}, \href
  {https://ui.adsabs.harvard.edu/abs/2008PASP..120..777T} {120, 777}

\bibitem[\protect\citeauthoryear{{Thum}, {Agudo}, {Molina}, {Casadio},
  {G{\'o}mez}, {Morris}, {Ramakrishnan}  \& {Sievers}}{{Thum}
  et~al.}{2018}]{Thum+2018}
{Thum} C.,  {Agudo} I.,  {Molina} S.~N.,  {Casadio} C.,  {G{\'o}mez} J.~L.,
  {Morris} D.,  {Ramakrishnan} V.,   {Sievers} A.,  2018, \mn@doi [\mnras]
  {10.1093/mnras/stx2436}, \href
  {https://ui.adsabs.harvard.edu/abs/2018MNRAS.473.2506T} {473, 2506}

\bibitem[\protect\citeauthoryear{{Tramacere}, {Giommi}, {Perri}, {Verrecchia}
  \& {Tosti}}{{Tramacere} et~al.}{2009}]{Tramacere+2009}
{Tramacere} A.,  {Giommi} P.,  {Perri} M.,  {Verrecchia} F.,   {Tosti} G.,
  2009, \mn@doi [\aap] {10.1051/0004-6361/200810865}, \href
  {https://ui.adsabs.harvard.edu/abs/2009A&A...501..879T} {501, 879}

\bibitem[\protect\citeauthoryear{{T{\"u}rler}, {Courvoisier}  \&
  {Paltani}}{{T{\"u}rler} et~al.}{2000}]{Turler+2000}
{T{\"u}rler} M.,  {Courvoisier} T.~J.~L.,   {Paltani} S.,  2000, \aap, \href
  {https://ui.adsabs.harvard.edu/abs/2000A&A...361..850T} {361, 850}

\bibitem[\protect\citeauthoryear{{Urry} \& {Padovani}}{{Urry} \&
  {Padovani}}{1995}]{Urry&Padovani+1995}
{Urry} C.~M.,  {Padovani} P.,  1995, \mn@doi [\pasp] {10.1086/133630}, \href
  {https://ui.adsabs.harvard.edu/abs/1995PASP..107..803U} {107, 803}

\bibitem[\protect\citeauthoryear{{Valtaoja}, {L{\"a}hteenm{\"a}ki},
  {Ter{\"a}sranta}  \& {Lainela}}{{Valtaoja} et~al.}{1999}]{Valtaoja+1999}
{Valtaoja} E.,  {L{\"a}hteenm{\"a}ki} A.,  {Ter{\"a}sranta} H.,   {Lainela} M.,
   1999, \mn@doi [\apjs] {10.1086/313170}, \href
  {https://ui.adsabs.harvard.edu/abs/1999ApJS..120...95V} {120, 95}

\bibitem[\protect\citeauthoryear{{Zacharias}, {B{\"o}ttcher}, {Jankowsky},
  {Lenain}, {Wagner}  \& {Wierzcholska}}{{Zacharias}
  et~al.}{2017}]{Zacharias+2017}
{Zacharias} M.,  {B{\"o}ttcher} M.,  {Jankowsky} F.,  {Lenain} J.~P.,  {Wagner}
  S.~J.,   {Wierzcholska} A.,  2017, \mn@doi [\apj] {10.3847/1538-4357/aa9bee},
  \href {https://ui.adsabs.harvard.edu/abs/2017ApJ...851...72Z} {851, 72}

\bibitem[\protect\citeauthoryear{{Zacharias}, {B{\"o}ttcher}, {Jankowsky},
  {Lenain}, {Wagner}  \& {Wierzcholska}}{{Zacharias}
  et~al.}{2019}]{Zacharias+2019}
{Zacharias} M.,  {B{\"o}ttcher} M.,  {Jankowsky} F.,  {Lenain} J.~P.,  {Wagner}
  S.~J.,   {Wierzcholska} A.,  2019, \mn@doi [\apj] {10.3847/1538-4357/aaf4f7},
  \href {https://ui.adsabs.harvard.edu/abs/2019ApJ...871...19Z} {871, 19}

\bibitem[\protect\citeauthoryear{{Zamaninasab}, {Clausen-Brown}, {Savolainen}
  \& {Tchekhovskoy}}{{Zamaninasab} et~al.}{2014}]{Zamaninasab+2014}
{Zamaninasab} M.,  {Clausen-Brown} E.,  {Savolainen} T.,   {Tchekhovskoy} A.,
  2014, \mn@doi [\nat] {10.1038/nature13399}, \href
  {https://ui.adsabs.harvard.edu/abs/2014Natur.510..126Z} {510, 126}

\bibitem[\protect\citeauthoryear{{Zheng}, {Yang}, {Zhang}  \& {Wang}}{{Zheng}
  et~al.}{2017}]{Zheng+2017}
{Zheng} Y.~G.,  {Yang} C.~Y.,  {Zhang} L.,   {Wang} J.~C.,  2017, \mn@doi
  [\apjs] {10.3847/1538-4365/228/1/1}, \href
  {https://ui.adsabs.harvard.edu/abs/2017ApJS..228....1Z} {228, 1}

\makeatother
\end{thebibliography}

\bsp	
\label{lastpage}
\end{document}